\documentclass[apj,numberedappendix]{emulateapj}
\usepackage{color}
\usepackage{xcolor}
\usepackage[linktocpage]{hyperref}
\usepackage{graphicx,amssymb,amsmath,amsthm,amsfonts,epsfig,bm}

\newcommand{\nn}{\nonumber}
\newcommand{\be}{\begin{equation}} 
\newcommand{\ee}{\end{equation}}
\newcommand{\beq}{\begin{eqnarray}}
\newcommand{\eeq}{\end{eqnarray}}

\usepackage{colortbl}

\newcommand{\tn}{\textnormal}

\begin{document}

\title{A new method to constrain neutron star structure 
from quasi-periodic oscillations}

\author{Andrea Maselli\altaffilmark{1} George Pappas\altaffilmark{2}, Paolo 
Pani\altaffilmark{1}, Leonardo Gualtieri\altaffilmark{1}, 
Sara Motta\altaffilmark{3}
Valeria Ferrari \altaffilmark{1}, and Luigi Stella\altaffilmark{4}}

\altaffiltext{1}{Dipartimento di Fisica, ``Sapienza'' Universit\`a di Roma \& Sezione 
INFN Roma1, Piazzale Aldo Moro 5, 00185, Roma, Italy.}
\altaffiltext{2}{Department of Physics, Aristotle University of Thessaloniki, 54124 Thessaloniki, Greece.}
\altaffiltext{3}{University of Oxford, Department of Physics, Astrophysics, Denys Wilkinson Building, 
Keble Road, OX1 3RH, Oxford, United Kingdom.}
\altaffiltext{4}{INAF-Osservatorio Astronomico di Roma, via Frascati 33, 00078, 
Monteporzio Catone, Roma, Italy.}


\begin{abstract}
\noindent We develop a new method to measure neutron star parameters 
and derive constraints on the equation of state of 
dense matter by fitting the frequencies of simultaneous Quasi 
Periodic Oscillation modes observed in the X-ray flux of 
accreting neutron stars in low mass X-ray binaries. To this 
aim we calculate the fundamental frequencies of geodesic 
motion around rotating neutron stars based on an accurate 
general-relativistic approximation  for their external spacetime.
Once the fundamental frequencies are related to the observed 
frequencies through a QPO model, they can be fit to the data 
to obtain estimates of the three parameters describing the 
spacetime, namely the neutron star mass, angular momentum 
and quadrupole moment. From these parameters we derive information on the neutron 
star structure and equation of state. We present a proof of 
principle of our method applied to pairs of kHz QPO frequencies 
observed from three systems (\texttt{4U1608-52}, \texttt{4U0614+09} 
and \texttt{4U1728-34}). We identify the kHz QPOs with the azimuthal 
and the periastron precession frequencies of matter orbiting 
the neutron star, and via our Bayesian inference technique we derive 
constraints on the neutrons stars' masses and radii. This 
method is applicable to other geodesic-frequency-based QPO models. 
\end{abstract}

\section{Introduction}

\noindent Neutron stars (NSs), the densest stable stellar objects in 
the Universe, provide key information on the properties 
of cold, supra-nuclear density matter, strong gravitational 
fields and a variety of astrophysical processes that take place 
in or around them. Observations of NSs
extend over the entire electromagnetic spectrum, from radio frequencies to 
gamma rays. The recent detection of gravitational waves (GWs) from two 
coalescing NSs has opened another, entirely different, observational window. 
Combinations of measurements of NS parameters such as mass, radius, 
moment of inertia, angular frequency, tidal deformability or quadrupole moments, 
which have already been obtained (or are within reach of present and future 
instrumentation), hold the potential to constrain the equation of state (EoS) of 
dense matter and the structure of NSs to high precision and accuracy (see e.g. 
\citealt{Lattimer:2006xb,Hinderer:2018mrj}).  To this aim different methods 
exploiting different diagnostics have been pursued: for instance X-ray based 
methods have been devised to infer the mass and radius of NSs from the 
evolution of radius expansion Type I X-ray bursts, broadened Fe 
K$\alpha$line profiles, fastest rotation frequencies, periodic modulations 
resulting from NS rotation (for a review see \citealt{Watts:2016uzu,Lattimer:2019eez}). 
NICER X-ray observations of a rotation-powered NS have 
recently provided $\sim 8-12$\% precise mass and radius measurements, 
based on the latter method \citep{Riley:2019yda,Miller:2019cac}. 
Constraints on the EoS of NSs have recently been obtained also from the GW 
signal of  GW170817 \citep{Abbott:2018exr}. Present limitations in the 
determination of the EoS of NS arise from data paucity and/or quality 
(insufficient sensitivity and signal to noise ratio especially), modelling 
uncertainties and control of systematics.

The fast quasi-periodic oscillations (QPOs) in the X-ray flux of  NS 
low mass X-ray binaries are among the observables that can yield 
measurements of NS mass and radius. These signals appear as narrow features in the power
spectra of the light curves of accreting NS and black holes. While their interpretation is still debated, QPOs are believed to be produced in the inner regions of the accretion disc 
surrounding the compact object. They often display different modes, some 
which are excited at the same time, and undergo correlated frequency changes 
(for a review see \citealt{2006csxs.book...39V}). Much of what is currently 
known about QPOs derives from observations with the 
Rossi X-Ray Timing Explorer  (RXTE, \citealt{Swank:1994zg}).  
RXTE detected a 
large number of low frequency QPOs (LF QPOs, observed below several 
tens of Hz) and enabled the discovery of high frequency QPOs (HF QPOs), 
often observed in pairs, with frequencies up to over a kHz in NS systems 
(the \textit{upper} and \textit{lower} kHz QPOs).

Following the realisation that kHz QPO 
frequencies are close to dynamical frequencies around compact objects \citep{vanderKlis1996,Strohmayer:1996zz}, 
the potential of QPOs for diagnosing matter motion in strong gravitational 
fields became apparent. This stimulated the development of models involving 
the innermost regions of accretion flows where general relativistic departures 
from Newtonian gravity are large (see e.g.\citealt{Belloni:2014iqa}). In a generic 
model that has been widely adopted since, the upper kHz QPO signal is directly 
associated to the azimuthal frequency of matter orbiting in the inner disk region, as \cite{1973SvA....16..941S} and others had suggested decades earlier). 
More complex {\it local} models aiming at interpreting also other QPO modes, the 
low frequency QPOs and the lower kHz QPO in particular, were soon proposed, 
building on the idea that QPOs are excited at a specific radius in the disk. 
Among these are the Relativistic Precession Model (RPM 
\citealt{Stella:1997tc,Stella:1998mq}) and the epicyclic resonance model (ERM 
\citealt{Abramowicz:2001bi,Kluzniak:2001ar,Kluzniak:2002bb, Fragile2016}). 
The frequencies of the signals predicted for different QPO modes consist of 
combinations of the fundamental frequencies  $\nu_{\phi}$, $\nu_{\theta}$ and 
$\nu_r$ of quasi-circular geodesics in the strong field regime. Also QPO models 
involving {\it global} oscillations of the inner disk, such as g- and p-modes 
\citep{1992ApJ...393..697N,1991ApJ...378..656N,Nowak:1996hg,Wagoner:2001uj,Kato:2004vs,kato:2012hq,Kato:2012mb} 
and corrugation and warping modes \citep{Markovic:1998qy,Armitage:1999aj,2001PASJ...53L..37K}
predict frequencies related to the frequencies of geodetic motion. 

As geodesics are determined by the spacetime metric, the above QPO models 
as well as models that extend them in different ways (see e.g. \citealt{Torok:2014eaa,2016A&A...586A.130S} 
and reference therein) hold the potential to probe the spacetime itself and measure 
key parameters of the compact object. In applications to black holes, the Kerr 
spacetime is generally used. The varying frequencies of two (and in one case three) QPO 
modes measured from a few black hole transients were successfully fit to the 
frequencies predicted by the RPM, and black hole mass and spin 
inferred \citep{Motta2014,2014MNRAS.439L..65M}. Similarly the $3:2$ frequency 
ratio in some black hole QPOs were exploited in the ERM to infer black hole 
parameters (e.g. \citealt{Torok2005}). The spacetime around rotating NSs is more complex, as it is characterised by mass, angular momentum, and higher 
order multipole moments. Analytical approximations to the fundamental 
frequencies of geodesic motion around NSs were used in early applications of the 
RPM (e.g. \citealt{Stella:1997tc}). Other studies adopted instead the numerically 
calculated spacetimes around rotating NS models, for selected EoSs 
\citep{1999ApJ...513..827M,Stella:1999sj}. 

\smallskip 

In this paper we introduce a general method to compute geodesic 
motion in the strong-field spacetimes generated by NSs, which builds on the 
finding that higher order multipole moments of NSs 
are related, to a good approximation, by some``three-hair relations'' 
\citep{Pappas:2013naa,Stein:2013ofa,Yagi:2014bxa,Yagi:2016bkt}. 
In this framework, fundamental frequencies of geodesic motion can be computed 
for any stationary and axisymmetric spacetime once the mass, angular momentum 
and quadrupole moment are specified \citep[for some additional applications, 
see][]{Maselli:2014fca,Maselli:2015tta}. These frequencies can then be used 
within QPO models to fit the observed QPO frequencies, test the assumptions of
any given model (such as, {\it e.g.}, the association of certain predicted 
frequencies with a given set of QPOs), and ultimately measure NS parameters 
that can be translated into constraints to the EoS.
An explicit form of the space-time metric in terms of mass, angular momentum, and higher order multipole moments has been derived by~\cite{Pappas:2016sye}\footnote{We note that our metric is a different, more accurate approach than the second order Hartle-Thorne 
spacetime which neglects multipole moments higher than 
the quadrupole, and is not valid for larger rotation rates.}. Such a metric enables to reproduce the orbital features of rotating NSs computed in fully relativistic numerical simulations up to the innermost stable circular orbit, with better than 1\% accuracy \citep{Pappas:2016sye}.
In this work we adopt the above metric, which enables to extend previous work on NS QPOs using analytic spacetimes \citep{Pappas:2012nt,Pappas:2012nv,Pappas:2015mba,Tsang:2015bty,Pappas:2016sye}.

Our aim is to measure/constrain the NS parameters by modelling the QPOs in the X-ray light curve of accreting NS binaries, making use of an accurate description of the NS surrounding spacetime. 
In this paper we present a first application of a new method to do so that  differs from similar methods previously proposed in the use of (i) an accurate metric to describe the spacetime of rapidly rotating NS, (ii) a Bayesian inference technique, which yields better constraints as the number of QPOs considered increases as it can take into account the information of all the QPOs consiered \textit{at the same time.} 
We use QPOs measured with RXTE from NS LMXBs with known rotation period (from 
oscillations observed during type-I X-ray bursts, see, e.g., 
\citealp{Watts2012}). We considered the sample of \cite{vanDoesburgh2017}, 
and selected three sources with a large number of QPO \textit{triplets} (i.e. 
two kHz QPOs and a LF QPO observed simultaneously), namely \texttt{4U1608-52},
\texttt{4U0614+09}, and \texttt{4U1728-34}.

Here we consider only the twin kHz QPOs; in a follow-up paper (Maselli et al., in prep) we will report on the results obtained by considering QPO triplets.
We follow the prescriptions of, e.g., the RPM, and we identify the upper and lower kHz QPOs with the azimuthal frequency $\nu_\phi$, 
and the periastron precession frequency, $\nu_\tn{per} = \nu_\phi-\nu_r$ 
(Stella \& Vietri 1999)\footnote{According to the RPM, a third frequency (historically known as \textit{horizontal branch oscillations}, see \citealp{vanderKlis1995a}) is associated with the nodal precession frequency, $\nu_\tn{nod}=\nu_\phi - \nu_\theta$  (Stella \& Vietri 1998; Stella et al 1999).}.
The two QPO signals (three, when a third QPO is considered)
are assumed to be generated at the same orbital radius. Correlated QPO frequency variations are thus the result of variations in the radius at which the QPOs are emitted (see also 
\citealp{Motta2014}). 

In this paper we shall use geometric units $G=c=1$. The mass will be 
expressed either in kilometres, or in solar masses $M_\odot=1.4768$ km.

\section{The metric for a rotating neutron star}\label{Sec:metric}

The metric around a rotating NS is expressed as a stationary and axisymmetric 
spacetime; it can be parametrised in terms of the first five relativistic multipole 
moments \citep{Geroch70I,Geroch70II,hansen:46,fodor:2252}, i.e., the mass $M$, 
the angular momentum $J$, the mass quadrupole $M_2$, the spin octupole $S_3$, 
and the mass hexadecapole $M_4$ \citep{Pappas:2016sye}. The line 
element for such a spacetime can be written as \citep{Papapetrou},
\be \label{Pap} ds^2=f^{-1}\left[ e^{2\zeta} \left( d\rho^2+dz^2 \right)+ \rho^2 d\varphi^2 \right]-f\left(dt-\omega d\varphi\right)^2\ ,
\ee
where ($\rho,z$) are the Weyl-Papapetrou coordinates 
and the metric components $f,\;\omega,$ and $\zeta$, 
shown in Appendix~\ref{app:metric}, are functions of 
the multipole moments and of the coordinates 
$(\rho,z)$\footnote{This metric is an approximate 
vacuum solution of Einstein's field equations, that 
is accurate up to $M_4$ in the moments and up to 
sixth order in $M/\sqrt{\rho^2+z^2}$.}.

To adjust the spacetime to the stellar structure of the central object, 
the right set of multipole moments must be specified. Recent work 
\citep{Pappas:2013naa,Yagi:2014bxa} has shown that  for NSs the first 
few relativistic multipole moments can be expressed as,
 \be\label{multipole}  M_2 = -\alpha j^2 M^3 , \quad
          S_3 = -\beta j^3  M^4, \quad
          M_4 =  \gamma j^4 M^5, 
     \ee
where $M$ is the mass and $j=J/M^2$ is the spin 
parameter, with $J$ being the angular 
momentum of the star. For NSs the coefficients $\alpha$, 
$\beta$, and $\gamma$ can be much larger than $1$, in contrast to Kerr black holes
(see \cite{DDGP2017arXiv} for a review). Furthermore 
it has been shown that for realistic EoSs, based on microphysical 
calculations, the higher order NS multipole moments (higher than $M_2$) can 
be expressed in terms of the quadrupole, angular 
momentum and mass (\citet{Pappas:2013naa,Stein:2013ofa,Yagi:2014bxa}; 
we refer the reader to \cite{Pappas:2016sye} for a 
detailed discussion). The spin octupole and 
the mass hexadecapole of a NS are related to the 
quadrupole by the relations\,\footnote{Eqs.~\eqref{uniRelS3}, \eqref{uniRelM4} 
are accurate up to $5\%$ for all state-of-the-art hadronic NS 
EoSs~\citep{Yagi:2014bxa,Pappas:2016sye}. In the case of quark stars, similar 
relations hold with slightly different values of the coefficients. }
\beq y_1&=&-0.36+1.48\, x^{0.65},\label{uniRelS3}\\
         y_2&=&-4.75+0.28\, x^{1.51}+5.52\, x^{0.22},\label{uniRelM4}\eeq
where $y_1=\sqrt[3]{-\bar{S}_3}=\sqrt[3]{\beta}$, $y_2=\sqrt[4]{\bar{M}_4}=\sqrt[4]{\gamma}$, 
$x=\sqrt{-\bar{M}_2}=\sqrt{\alpha}$, and $\bar{M}_n=\frac{M_n}{j^nM^{n+1}}$, $\bar{S}_n=\frac{S_n}{j^nM^{n+1}}$ 
are the (dimensionless) reduced moments. For NSs $\alpha$ varies  in the range 
between $1.5$ and $10$ for masses between $1M_{\odot}$ and up to the maximum 
mass (which is different for each EoS), where smaller values correspond to larger masses.

Therefore, 
the description of the spacetime and of the various 
geodesic frequencies will only depend on three 
parameters: the mass $M$ (that we express in units 
of kilometers in $G=c=1$ units), the dimensionless 
spin parameter $j$, and the dimensionless reduced 
quadrupole $\alpha\equiv-\bar M_2$. This is especially 
relevant for our analysis, as it reduces the number 
of parameters to be constrained. 

By relating QPO frequencies to geodesic frequencies 
through the adoption of a given model (the RPM in our 
present case), the mass $M$, reduced spin $j$ and 
reduced quadrupole moment $\alpha$ can be measured as
independent parameters determining the characteristics 
of the spacetime. These three parameters, if measured precisely, 
provide constraints on the NS EoS. This {\it top-down} approach 
involves no {\it a priori} assumption on the EoS and contrasts with 
other studies in which geodesic frequencies are compared 
to QPO frequencies based on the spacetime calculated for 
individual EoS and specific values of $M$ and $j$ 
(see e.g. Morsink \& Stella 1999; Torok et al. 2016)

The spacetime considered in this work is 
equivalent to that introduced in \citep{Pappas:2016sye} 
and valid for all rotation rates (see 
Appendix~\ref{app:metric} for technical details).
Our metric is different and more accurate than slowly-rotating 
approaches, as the second order Hartle-Thorne spacetime (see 
\citet{Urbancova:2019btk} for recent and detailed analyses 
using this approach).
It extends previous work on NS QPOs based on analytic spacetimes 
\citep{Pappas:2012nt,Pappas:2012nv,Pappas:2015mba,Tsang:2015bty,Pappas:2016sye}.
The accuracy of the epicyclic frequencies used in this paper 
has been tested against fully numerical solutions showing an agreement 
better than 99\% down to the innermost stable circular 
orbit \citep{Pappas:2016sye}, which outperforms the approaches described above.

\section{Numerical analysis}
\label{Sec:numerical}

In order to test our method with the QPO frequencies  
from the sources in our sample we adopted a Bayesian approach. 
For a given set of $n$ observations $\vec{O}$ we 
wish  to determine the posterior probability distribution 
of the system's parameters 
$\vec{\theta}=(r_{i=1,\ldots n},M,j,\alpha)$, i.e.
\begin{equation}
{\cal P}(\vec{\theta}\vert\vec{O})\propto {\cal P}_0(\vec{\theta}){\cal L}(\vec{O}\vert\vec{\theta})\ ,\label{posterior}
\end{equation}
where ${\cal P}_0(\vec{\theta})$ represents the prior 
information on the parameters. ${\cal L}(\vec{O}\vert\vec{\theta})$ 
is the likelihood function, which we assume proportional 
to a chi-square variable, ${\cal L}\propto e^{-\frac{1}{2}\chi^2}$, 
given by:  
\begin{equation}
\chi^2=
\sum_{i=1}^{N_\tn{obs}}\left(\frac{\Delta_\phi^2}{\sigma^2_{\nu_{\phi}}}+\frac{\Delta_{\rm per}^2}{\sigma^2_{\nu_{\rm per}}}\right)
           \,,\label{chi2}
\end{equation}
%
where $N_\tn{obs}\ge N_\tn{min}$, $\Delta_k\equiv \nu_k^{\rm obs}-\nu_k(\vec{\theta})$, and $\nu_j$ can be either 
the azimuthal or the periastron precession frequency.

Since we use here only pairs of kHz QPO frequencies ("doublets"), each source provides 
$2N_\tn{obs}$ frequencies, which are used to determine $3+N_\tn{obs}$ unknown parameters, 
{\it i.e.} the NS parameters $(M,j,\alpha)$ and the circumferential radii $r_i$ where 
each is produced. Thus, we need at least $N_\tn{min}=3$ doublets to characterise each source. 
As the errors of the QPO observed frequencies are in general asymmetric (see Table~\ref{table_freqs}), 
i.e. $\nu_j=\nu_j^\tn{obs}\pm \sigma^{(\pm)}_{\nu_j}$, with $\sigma^{(+)}_{\nu_j}\neq \sigma^{(-)}_{\nu_j}$, 
for the sake of simplicity we compute the chi-square functions of Eq.~\eqref{chi2} by using their
$\sigma_{\nu_j}=[\sigma^{(+)}_{\nu_j}+\sigma^{(-)}_{\nu_j}]/2$. 
This has a negligible effect on inferred values of the source parameters. 
Moreover, given the accuracy of our metric, the relative difference
between the actual frequencies and those computed using 
\eqref{Pap} is subdominant with respect to the observational 
errors, and therefore it will be neglected in our analysis \citep{Pappas:2016sye}.

We sample the posterior distribution \eqref{posterior} 
using a Markov Chain Monte Carlo~(MCMC) approach, based 
on the Metropolis-Hastings algorithm \citep{Gilks:1996}. 
The random jump within the parameter space is chosen 
according to a multivariate Gaussian distribution, whose 
covariance matrix is continuously updated through a Gaussian 
adaptation scheme \citep{5586491}, which increases the 
mixing of the chains and boosts the convergence to the 
target distribution. For each set of data, we run four  
independent chains of $2\times10^6$ samples, generally 
discarding the first $10\%$ of the simulation as a burn 
in. The convergence of the processes is then assessed by 
a standard Rubin test. 

We consider flat prior distributions for all the parameters 
within the following ranges: $M\in[0.7,3]M_\odot$, $j\in [0,0.7]$, 
$\alpha\in[1.5,15]$, $r_i\in[R_\tn{NS},15M]$. Note that 
the prior on $\chi$ allows only co-rotating orbital motion.
We also set the prior on $r_i$ such that it is always larger 
than the stellar equatorial circumferential radius $R_\tn{NS}$. The 
latter can be expressed with very good accuracy (a 
few percent) as a function of $M,j$, and $\alpha$:
\begin{equation}
R_\tn{NS}/M=\sum_{i=0}^3[{\cal B}_i j^i+{\cal A}_i j^i \alpha^{{\cal N}_1/2}+{\cal C}_i j^i \alpha^{{\cal N}_2/2}]\ ,\label{radiusfit}
\end{equation}
were the numerical coefficients 
$({\cal B}_i,{\cal A}_i,{\cal C}_i,{\cal N}_1,{\cal N}_2)$ 
are listed in Table~\ref{table_fit}.

\begin{table}[ht]
\centering
\begin{tabular}{cccc}
\hline
\hline
&   $\ell=0$ & $\ell=1$ & $\ell=2$\\
$A_{\ell}$ & 0.00927584 & -0.0252801 & 0.0497335\\
$B_{\ell}$ & -0.358824 & 3.15892 & -5.30171\\
$C_{\ell}$ & 2.94923 & -3.20369 & 6.02522\\
\hline
\hline
\end{tabular}
\caption{Numerical coefficients of the empirical 
relation \eqref{radiusfit}, which provides the NS circumferential radius 
as a function of the dimensionless spin parameter and of the 
stellar quadrupole. The best fit values for the exponents 
${\cal N}_{1,2}$ are given by ${\cal N}_1=4.12566$ and 
${\cal N}_2=0.996284$ \citep{Pappas:2015mba}.}\label{table_fit}
\end{table}

Once the  ${\cal P}(\vec{\theta}\vert\vec{O})$ is sampled 
by the MCMC, we derive the probability distribution of the source parameters, by marginalising the joint 
posterior distribution over the emission radii:
\begin{equation}
{\cal P}(M,\chi,\alpha)=\int {\cal P}(\vec{\theta}\vert\vec{O}) dr_1\ldots d r_n\ .
\end{equation}
In the following sections we apply this analysis to the kHz QPO of
\texttt{4U1608-52}, \texttt{4U0614+09}, \texttt{4U1728-34}, 
and compute 
${\cal P}(M,\chi,\alpha)$ using various combinations of 
the doublets for each source, (see list 
in Table~\ref{table_freqs}). The latter step is a crucial  
to our analysis as it provides a self-consistency test for 
the applicability of the RPM in conjunction with the geodesic frequencies
calculated in our method. If the assumptions of the method are correct, 
different sets of doublets for a given source 
must necessarily provide probability distributions for masses, spins and quadrupole 
moments which are consistent with each other to within the
uncertainties. On the contrary, an inconsistency of the 
different probability distributions would signal problems 
with the adopted geodesic and/or QPO model.

\section{Results}\label{Sec:results}

\begin{figure*}[!htbp]
\centering
\includegraphics[width=0.25\textwidth]{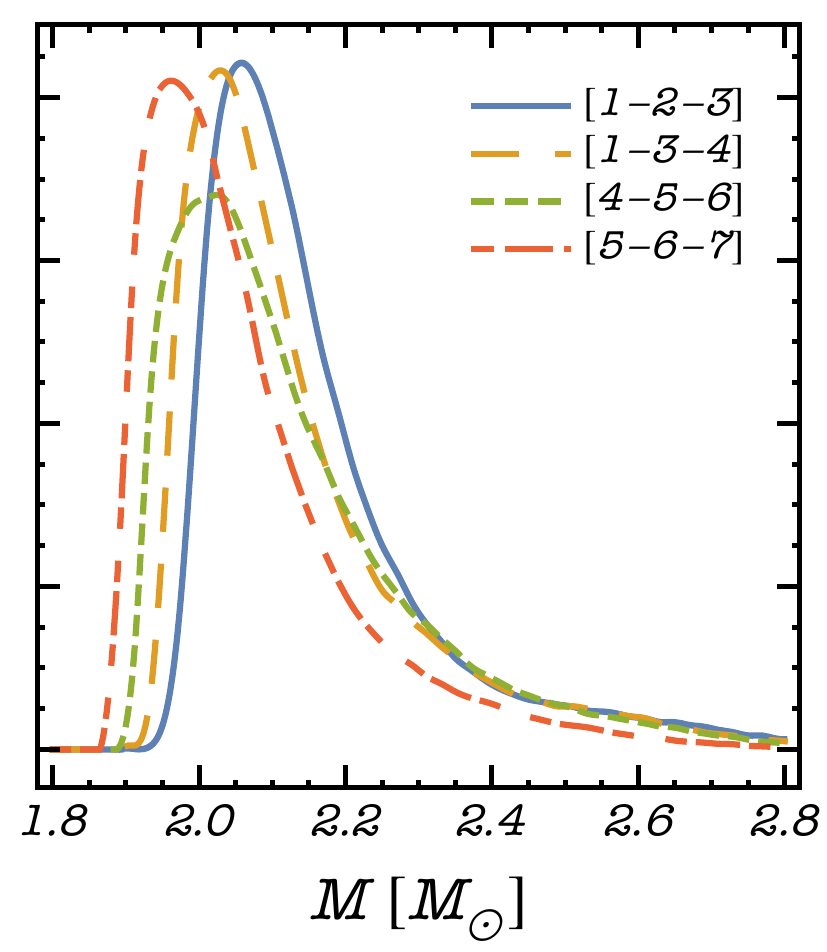}
\includegraphics[width=0.25\textwidth]{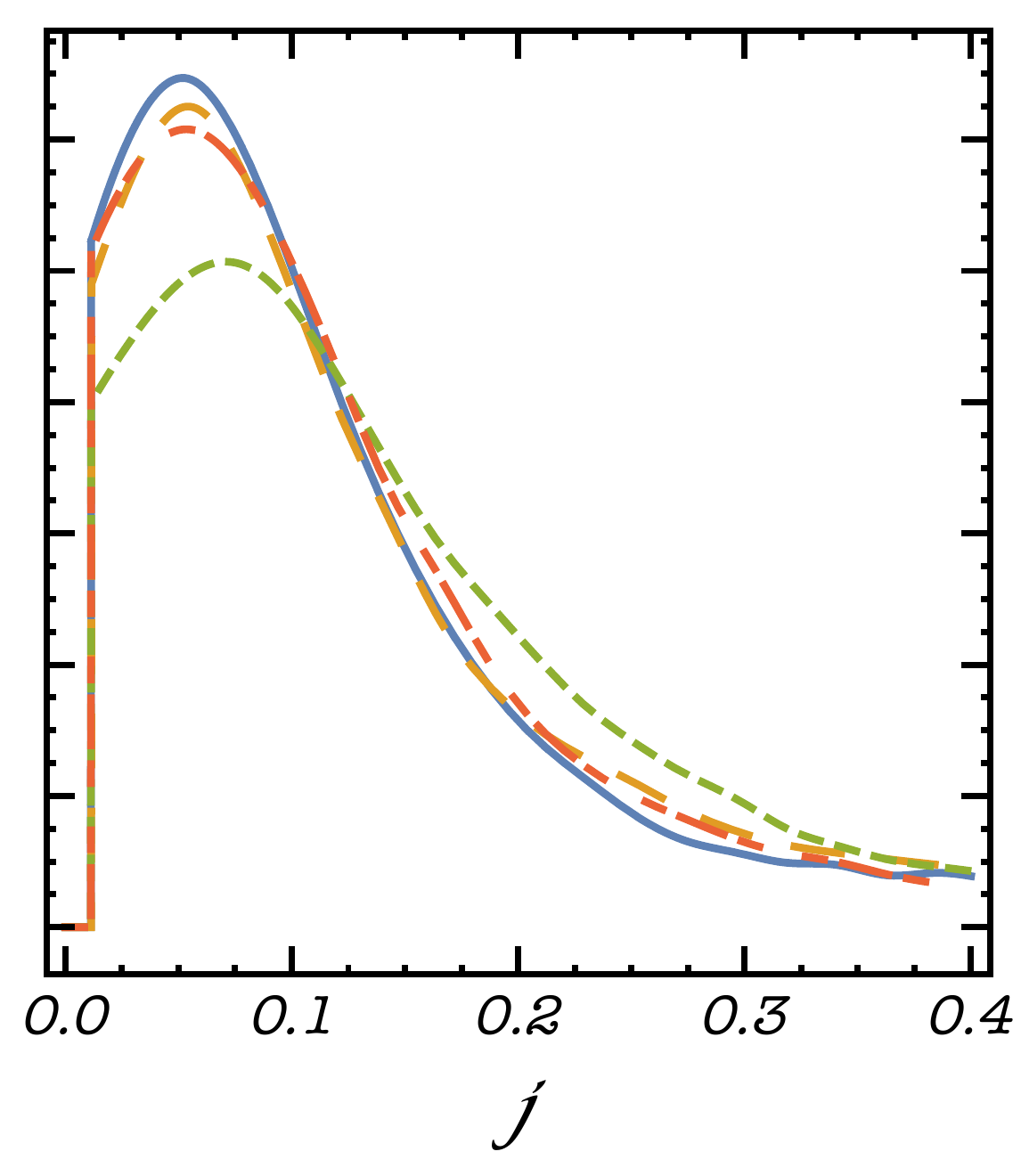}
\includegraphics[width=0.25\textwidth]{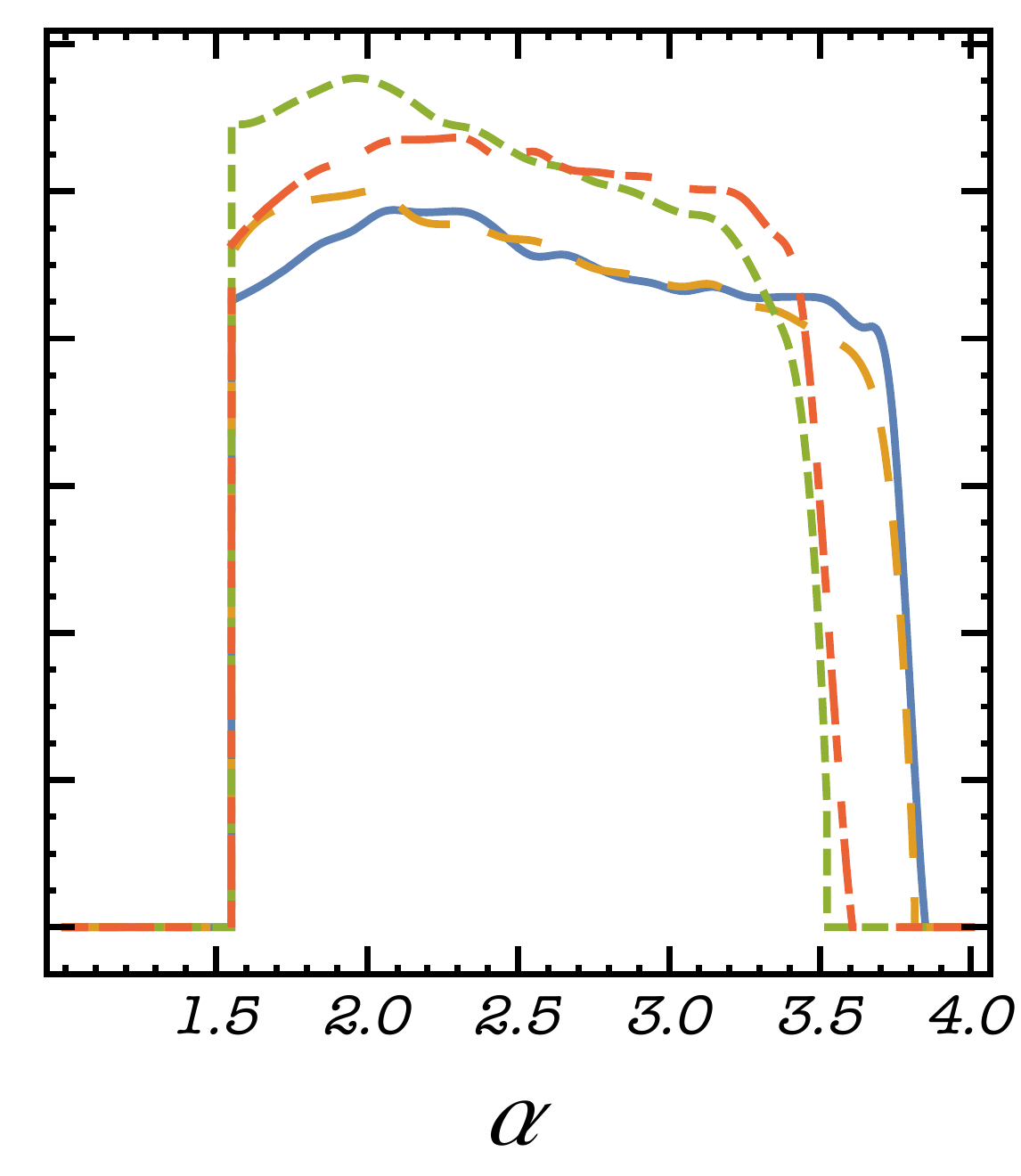}
\caption{Marginalised posterior probabilities of $(M,j,\alpha)$ 
for \texttt{4U1608-52} derived by using different groupings of 
three QPO doublets 
drawn from the observed frequencies $(\nu_\phi, \nu_\tn{per})$  in
Table.~\ref{table_freqs}.}\label{fig:doub4U1608-52} 
\end{figure*}

Since we analysed kHz QPOs doublets, {\it i.e.} the QPOs that in the 
RPM correspond to the periastron and the azimuthal 
frequencies, the MCMC requires for each source at least 
three sets of QPO frequencies, each containing the QPO doublet 
relevant for the model,  i.e. $(\nu_\phi,\nu_\tn{per})_{i=1,2,3}$, 
to be solved in terms of $(M,\alpha,j)$ and three 
emission radii $(r_1,r_2,r_3)$. 
Therefore, for each source, we randomly selected sub-sets of three doublets from the available data. Figure~\ref{fig:doub4U1608-52} shows the posterior probabilities obtained for different sets of  doublets, while the associated numerical values are reported in Table~\ref{tableparams}.
As a representative case, we describe in detail the results we obtained for 
\texttt{4U1608-52}. 

\begin{figure}[!htbp]
\centering
\includegraphics[width=0.48\textwidth]{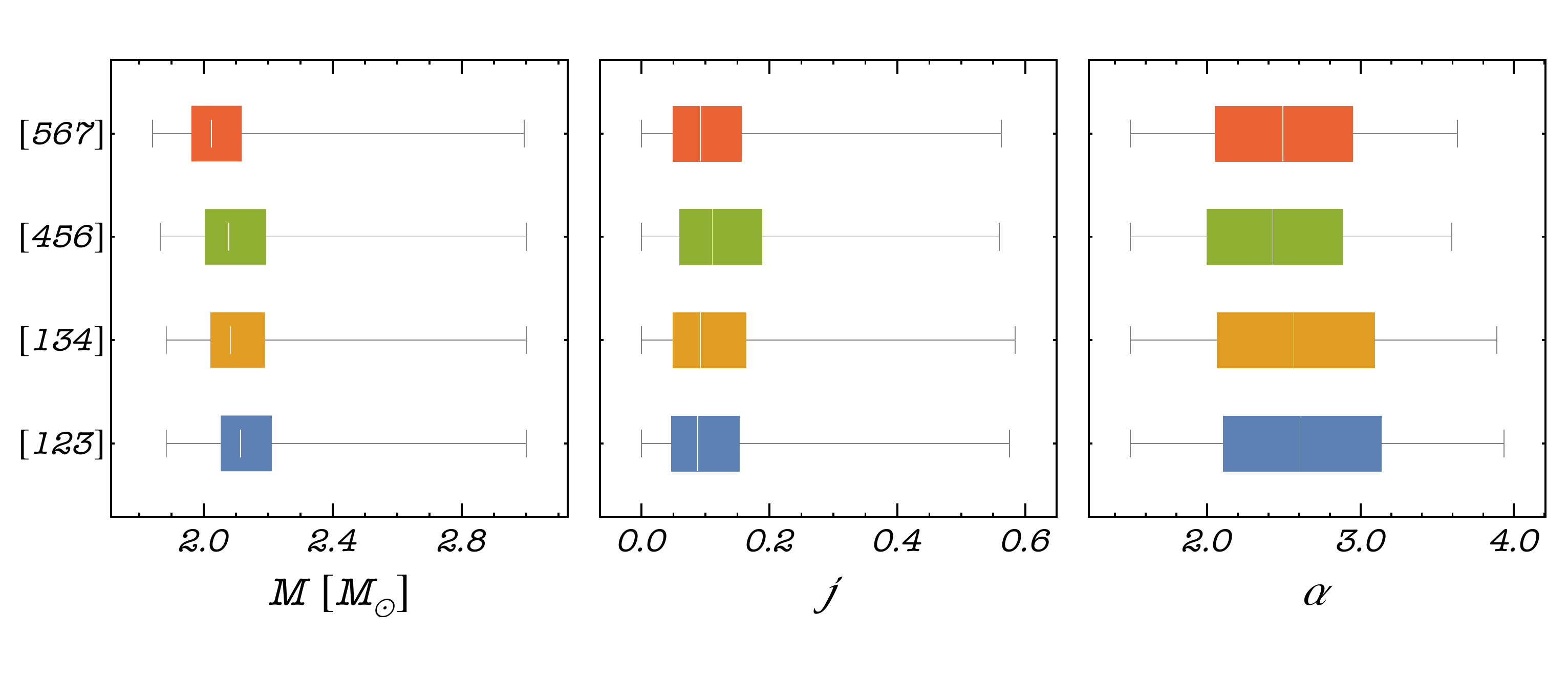}
\caption{Box and whiskers plots for $(M,j,\alpha)$, 
corresponding to the probability distributions shown in Fig.~\ref{fig:doub4U1608-52}. 
White vertical lines in each coloured box mark the median 
of the parameters. The edges of the box identify the upper 
and lower quartiles, while the ends of the whiskers yield 
the maximum and minimum inferred values.} 
\label{fig:whiskers_4U1608-52} 
\end{figure}

The posterior distributions obtained by using different datasets are in good agreement with each other, and the different parameter distributions are all remarkably consistent (even though in some cases shifts between the peaks of the different marginal distributions are apparent). However the quadrupole 
moment of the star remains essentially unconstrained, its probability distributions being almost flat 
between $\alpha\simeq 1.5$ and $\alpha\simeq4$. 
This is to be expected for two reasons: firstly, the quadrupole moment gives a sub-leading contribution to the spacetime metric relative 
to mass and spin; and secondly, the effect of the quadrupole moment is largest on the  $\nu_\tn{nod}$  frequency - which we do not consider in this work - while $\nu_\phi$ and $\nu_\tn{per}$ 
are only weakly affected by its variation. 
However, it should be noted that adopting the approach described here with a larger set of QPO doublets would allow the quadrupole moment to be better constrained.
The box plot shown in Fig.~\ref{fig:whiskers_4U1608-52} 
supports the use of our approach together with the RPM model: it is
clearly seen that the median of the distributions are all 
consistent with each other, and the interquartile ranges 
overlap with good accuracy. 

\begin{figure}[!htbp]
\centering
\includegraphics[width=0.47\textwidth]{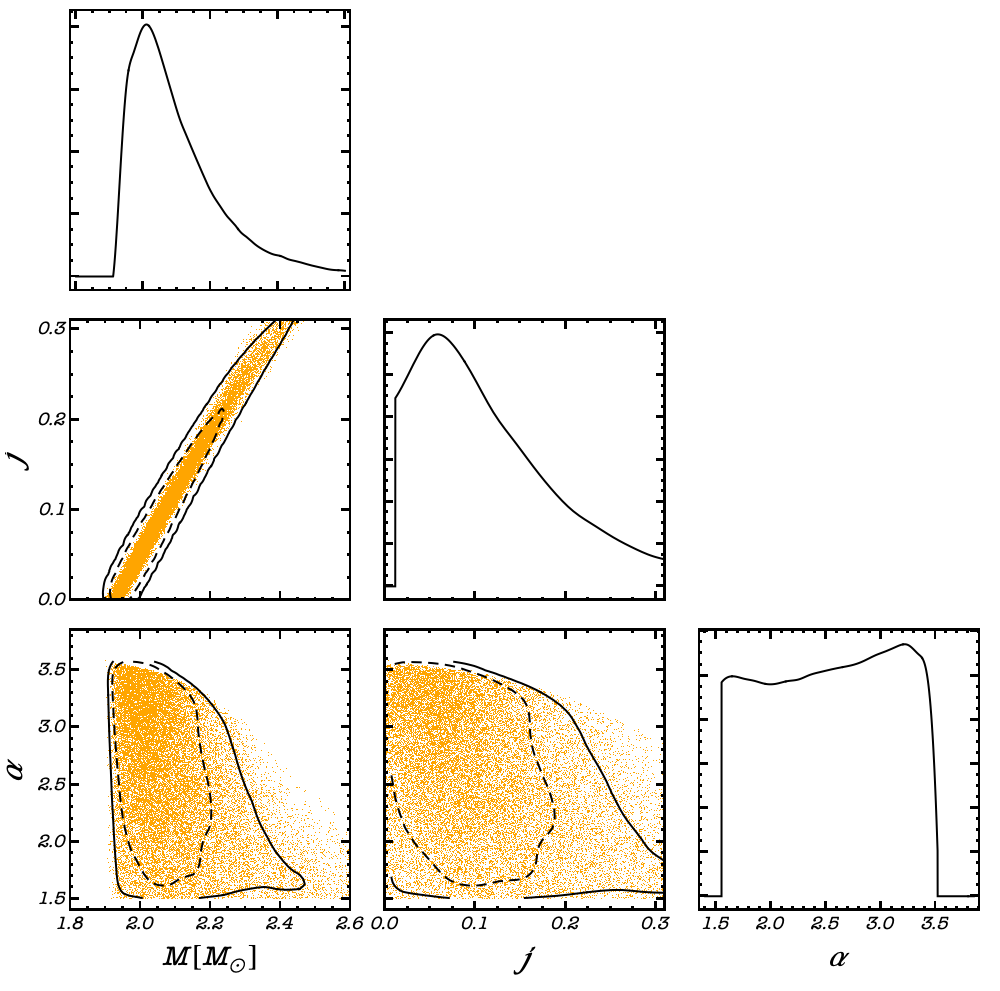}
\caption{Triangle plot for the posterior of the  
parameters of \texttt{4U1608-52}. Diagonal and off-diagonal 
panels refer to marginalised and 2D joint posterior 
distributions, respectively. Dashed and solid curves 
identify contours at 68\% and 90\% confidence intervals, 
while colored dots represent the actual points sampled 
by the MCMC.}
\label{fig:alldoublets_4U1608-52} 
\end{figure}

Motivated by the above results, we performed the same Bayesian analysis by progressively increasing the number of doublets. 
We found that in all cases the posterior distributions are consistent and yield acceptable parameters. 
We show the results of these steps in the triangle plot in 
Fig.~\ref{fig:alldoublets_4U1608-52}. The diagonal and 
off-diagonal panels show the marginalised and the 2D joint 
distributions of $M$, $j$, and $\alpha$, respectively. 
The mass is the parameter that we determine with the highest precision:
at $90\%$ confidence level\footnote{Following 
\citep{Abbott:2018wiz} for parameters bounded by the prior, 
as the quadrupole, we quote the $90\%$ one-sided upper 
limit, while for all the other parameters we quote the $90\%$ 
highest-posterior density interval.} we find $M\in[1.92,2.32]M_\odot$, 
with a median of $M=2.07M_\odot$. 

Our  analysis also provides a poorly constrained spin parameter with median $j= 0.1$
within the $90\%$ interval $j\in[0,0.26]$.
The 2D distributions of Fig.~\ref{fig:alldoublets_4U1608-52} show some degree of positive 
correlation between of $j$  and $M$. 
The inclusion of all datasets does not modify significantly 
the constraints on $\alpha$, for which we still obtain a flat posterior 
distribution within $\alpha\in[1.50,3.30]$ at $90\%$ 
(see the bottom row of Fig.~\ref{fig:alldoublets_4U1608-52}).  
In fact the values of the quadrupole  sampled by the Monte-Carlo simulation 
are essentially degenerate with respect to 
$M$ and $j$. As noted above, this result is somewhat 
expected as the azimuthal and periastron precession frequencies depend 
weakly on $\alpha$. 

\begin{figure}[!htbp]
\centering
\includegraphics[width=0.25\textwidth]{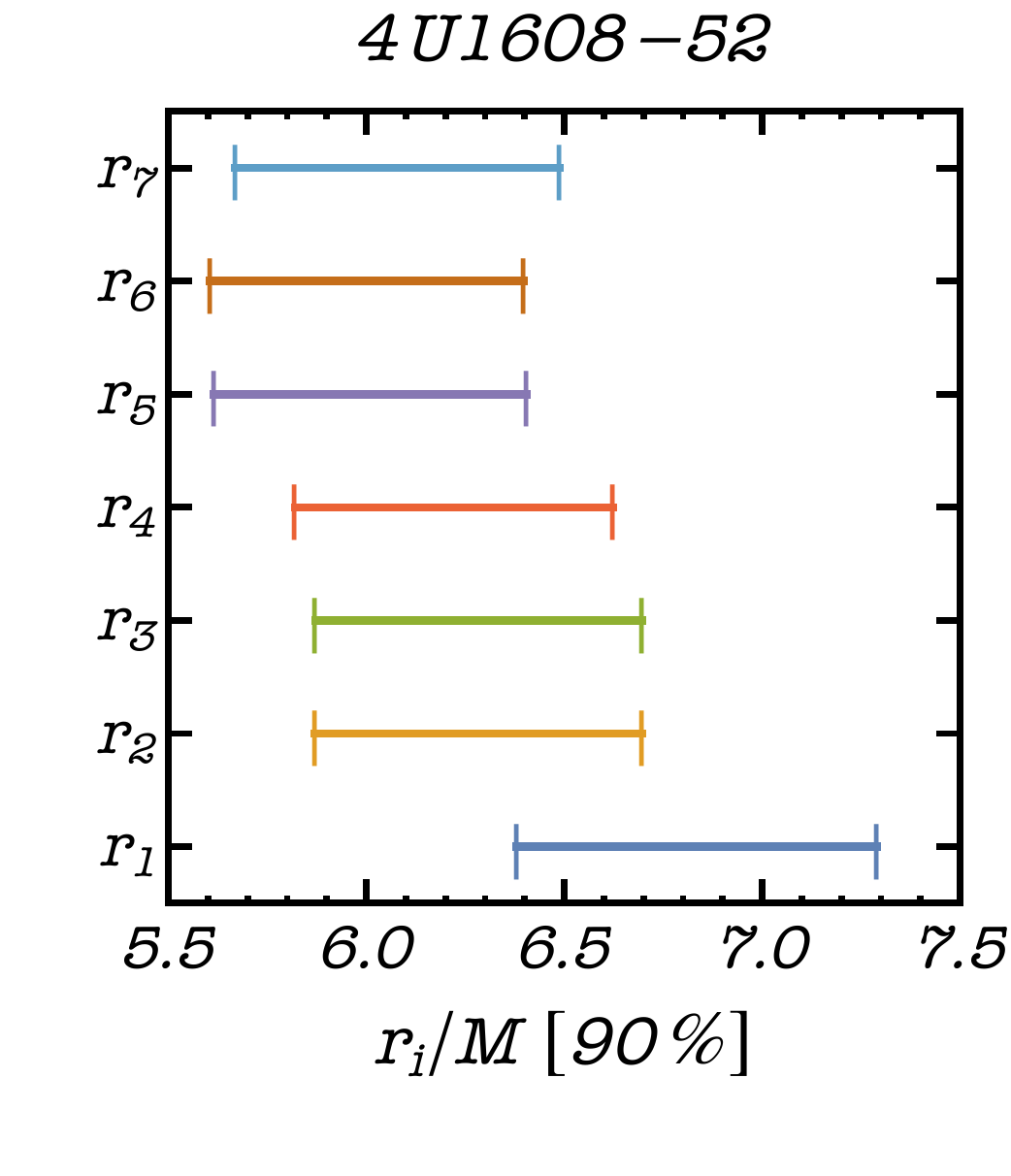}
\caption{90\% credible interval for the posterior 
distribution of the QPO radius  for the doublets of \texttt{4U1608-52}.}
\label{fig:radii_4U1608-52} 
\end{figure}

The MCMC also leads to constraints on the orbital radius associated with each doublet 
in the analysis. 
We find values
$r_i/M\lesssim 7$ in all cases. 
Figure~\ref{fig:radii_4U1608-52} shows the 
 the $90\%$ confidence intervals for the 
7 values. 
As noted in applications of simple geodesic models (e.g. Miller et al. 1998)
upper limits on the QPO radius derived the from the highest observed QPO 
frequencies provide an upper bound on the stellar radius. 
Requiring that the oscillations is generated within the accretion disc at orbital distances 
larger than $R_\tn{NS}$ yields the limit $R_\tn{NS}\lesssim 6.4 M$ 
at 90\% confidence level from the highest frequency Doublet of \texttt{4U1608-52} 
(see $r_6$ in Fig.~\ref{fig:radii_4U1608-52}).

\begin{figure}[!htbp]
\centering
\includegraphics[width=0.44\textwidth]{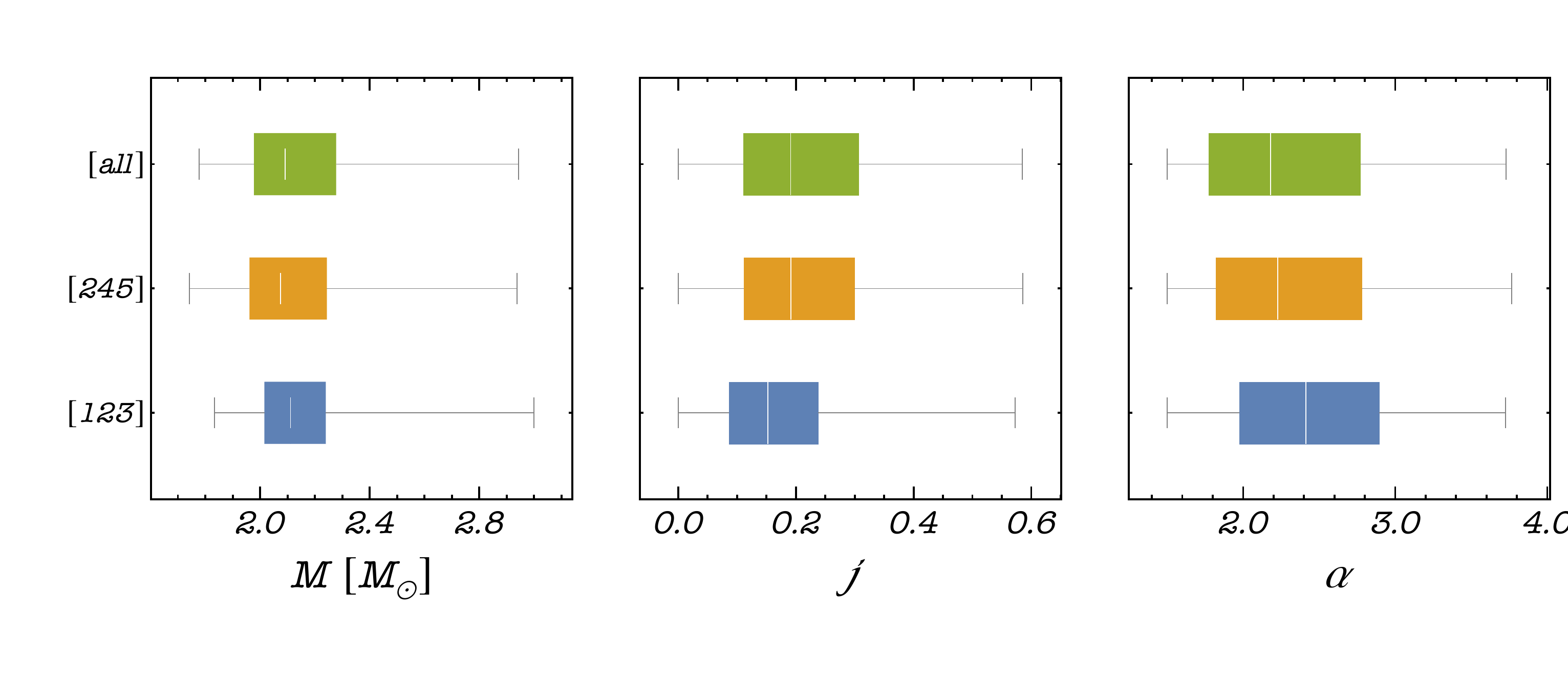}
\includegraphics[width=0.44\textwidth]{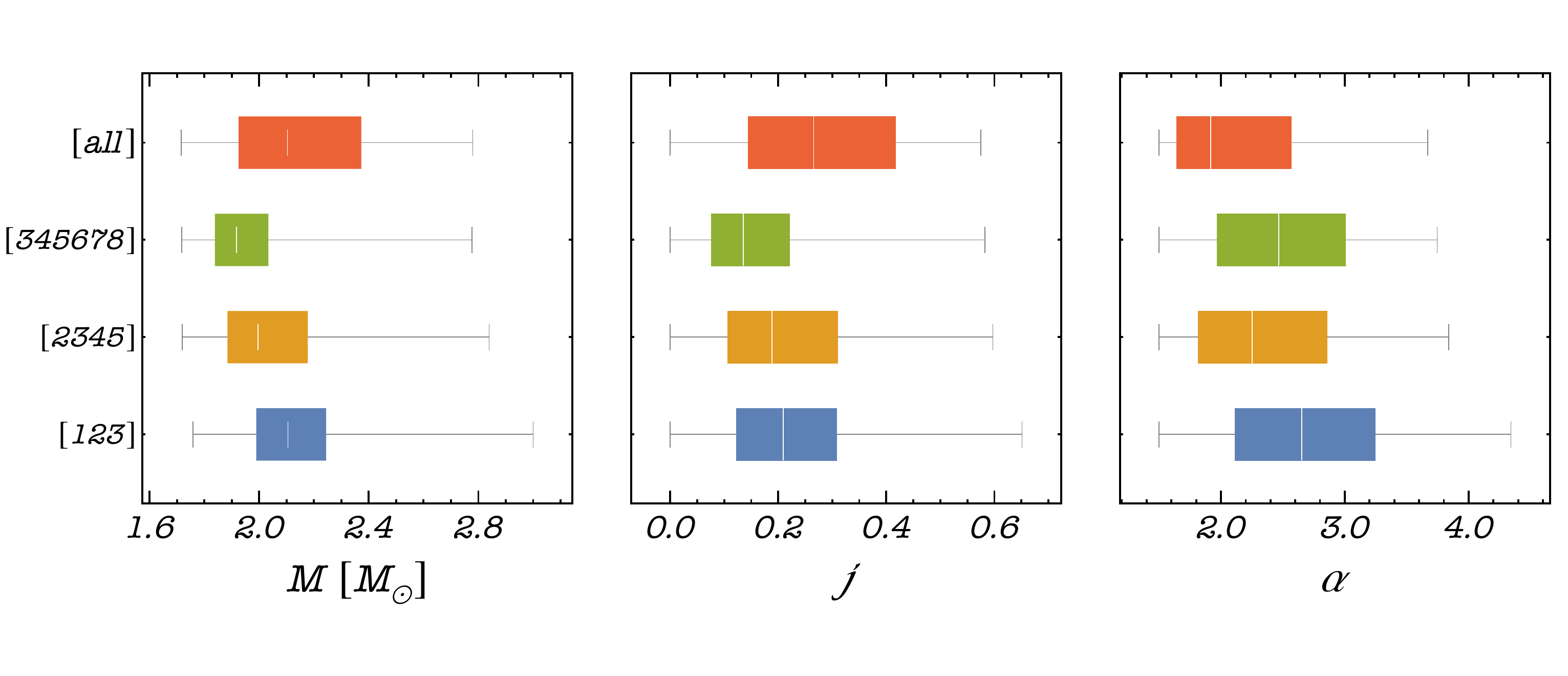}
\caption{Same as in Fig.~\eqref{fig:whiskers_4U1608-52} 
but for \texttt{4U0614+09} (top) and 
\texttt{4U1728-34} (bottom).}
\label{fig:wiskers2} 
\end{figure}

\begin{figure}[!htbp]
\centering
\includegraphics[width=0.21\textwidth]{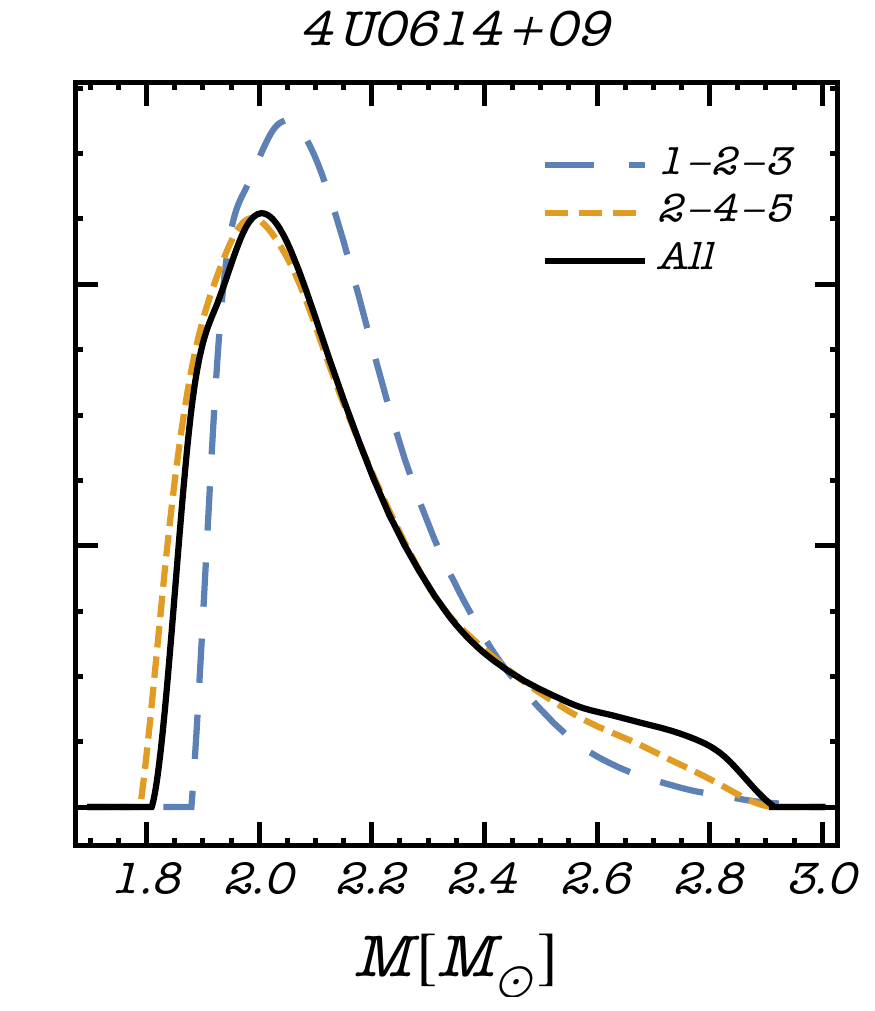}
\includegraphics[width=0.21\textwidth]{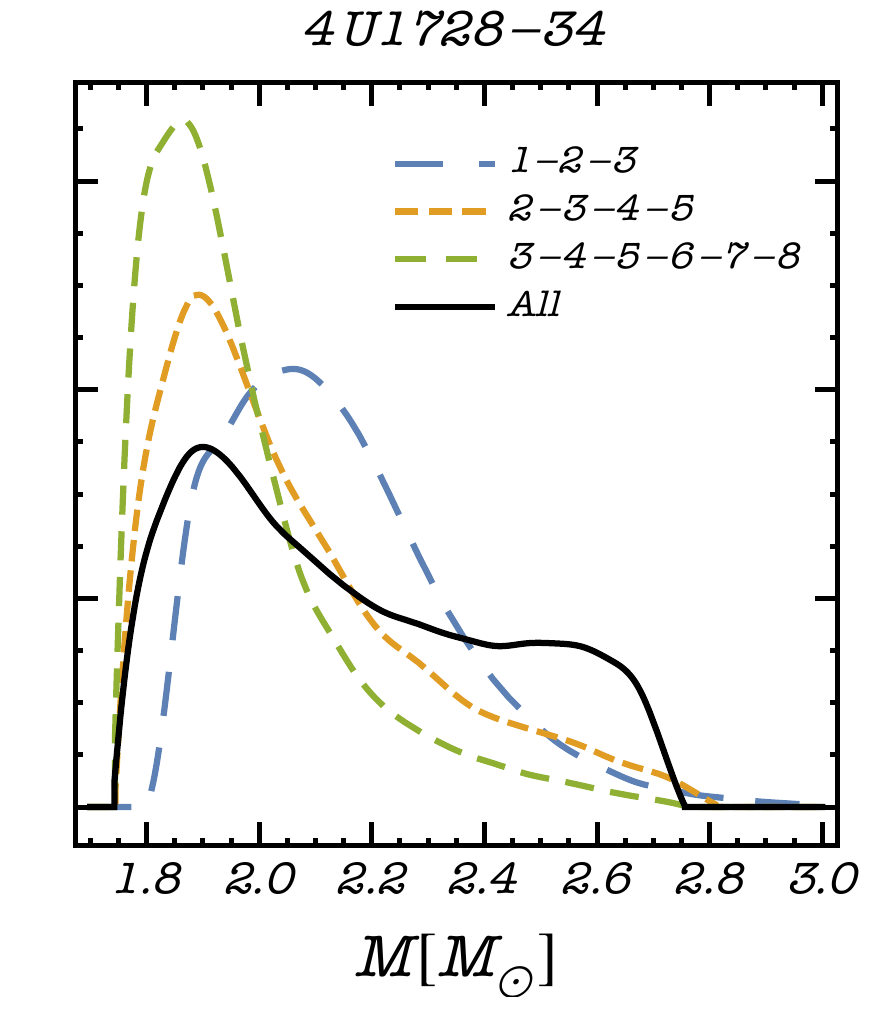}
\caption{Posterior distributions for the mass of the 
binary systems \texttt{4U0614+09} and \texttt{4U1728-34}. 
Dashed curves refer to different grouping of kHz QPO doublets, whereas 
the solid black lines are obtained by 
fitting all doublets for each source.}
\label{fig:MR_4U0614+09} 
\end{figure}

The analysis of the kHz QPOs of \texttt{4U0614+09} and \texttt{4U1728-34}
yields similar results to that of the \texttt{4U1608-52} kHz QPOs. 
Figure~\ref{fig:wiskers2} shows that also for these systems 
there exists a general agreement of the inferences  
from different subsets of doublets, with
\texttt{4U1728-34} displaying the largest spread in the posterior distributions 
of the parameters when the full set of doublet is taken into account. 
\texttt{4U0614+09} and \texttt{4U1728-34} are also characterised 
by a mass distribution which peaks around $\sim 2M_\odot$. 
The posteriors of $M$ are shown in 
Fig.~\ref{fig:MR_4U0614+09}; different
groupings of doublets appear to give consistent results for each system.
and converge to a similar distribution. 
These are fairly large mass values compared to those typically derived for, e.g., isolated pulsars \citep{Ozel2012}, but we note that other studies recovered similarly large NS masses (e.g., \cite{Stella:1998mq,Torok:2014eaa}). The quadrupole moments inferred for 
these two systems would suggest more compact objects (i.e, with larger values of $M/R$) with faster rotation rates with respect to \texttt{4U1608-52}, although we stress that the 
inferred spin parameters are not very well constrained. The
median of all parameters for each sources, 
together with $90\%$ uncertainties, are reported in 
Table~\ref{tableparams}.

\begin{table}[ht]
	\renewcommand*{\arraystretch}{1.2}
  \centering
     \begin{tabular}{ccccccc}
   \hline
 source && $M/M_\odot$ && $j$ && $\alpha$ \\
 \hline
\texttt{4U1608-52}  && $2.07_{-0.15}^{+0.25}$ && $0.10_{-0.10}^{+0.16}$ && $2.56^{+0.74}_{-1.06}$ \\
\texttt{4U0614+09}  && $2.10_{-0.27}^{+0.45}$ && $0.20_{-0.20}^{+0.24}$ && $2.22^{+1.04}_{-0.68}$ \\
\texttt{4U1728-34}  && $2.11_{-0.34}^{+0.47}$ && $0.27^{+0.24}_{-0.27}$ && $2.00^{+1.18}_{-0.46}$ \\
   \hline
   \hline
  \end{tabular}  
  \caption{Median and 90\% values for the parameters of the sources analysed.}\label{tableparams}
\end{table}

\begin{figure}[!htbp]
\centering
\includegraphics[width=0.4\textwidth]{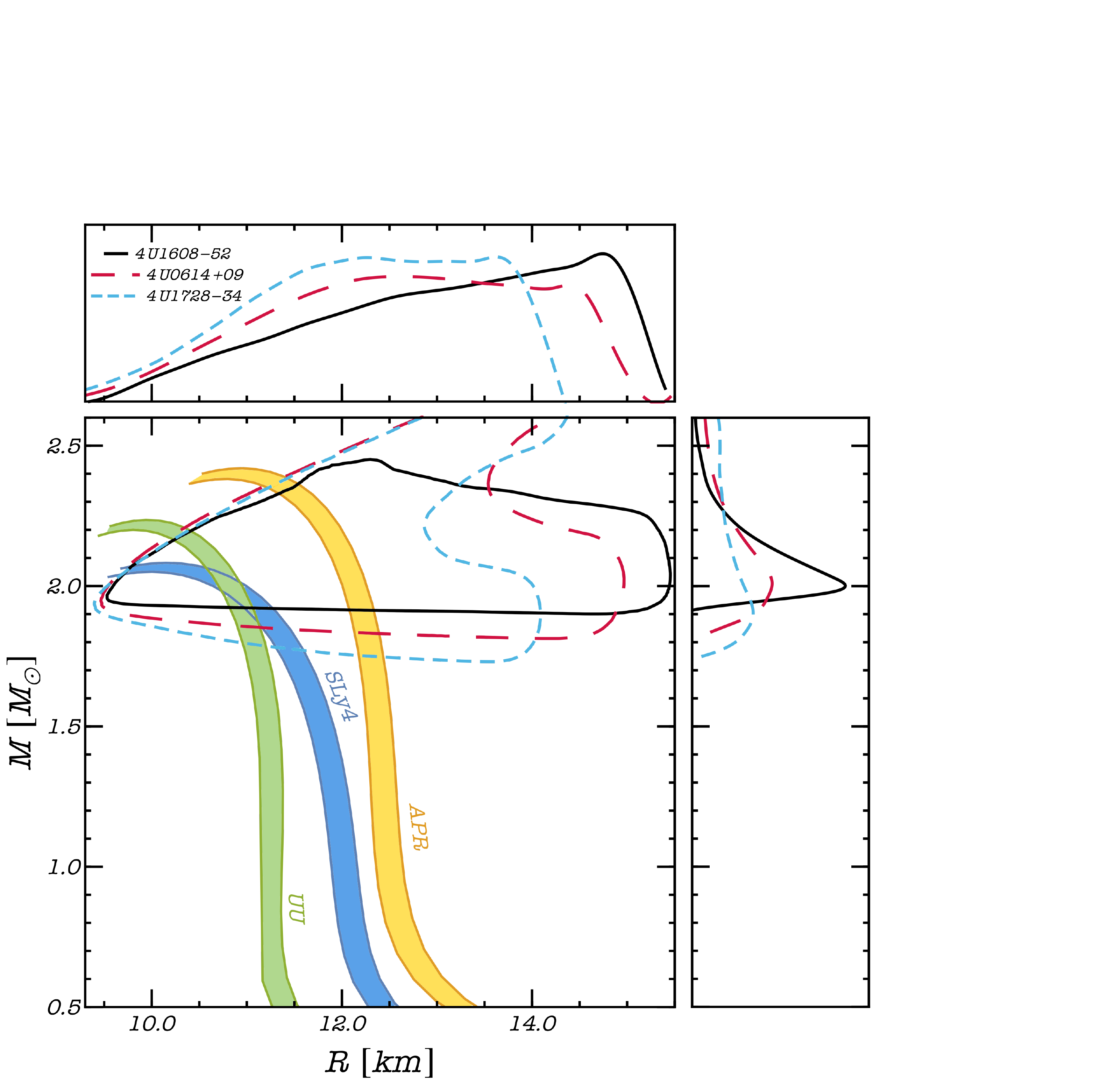}
\caption{90\% 2D credible interval and marginalised 
distributions of the mass and radius for the 3 sources 
analysed. Colored curves are from  
of a few EoS whose $M$-$R$ relation 
is compatible with current constraints from X-ray observations
and the gravitational-wave event GW170817.
The left and right edges of each colored band correspond to stellar 
configurations for which the spins are assumed to be 
zero and equal to the median of the values inferred by our MCMC analysis, respectively.}
\label{fig:massradius}
\end{figure}

The values of $(M,j,\alpha)$ provide an estimate for the 
NS equatorial radius through \eqref{radiusfit}. 
We used the values we determined from the 
analysis of the full set of doublets of \texttt{4U1608-52}, 
\texttt{4U0614+09}, and \texttt{4U1728-34} to build 
the joint 2D distribution of ${\cal P}(M,R_\tn{NS})$. 
The contour plot of Fig.~\ref{fig:massradius} shows 
the $90\%$ region of the mass-radius posteriors of 
the three sources.  Fig.~\ref{fig:massradius} shows also
the expected mass-radius relations  for three  hadronic EoS (APR 
\citep{Akmal:1998cf}, SLy4 \citep{Douchin:2001sv}, and 
UU \citep{PhysRevC.38.1010}), are all consistent
with current constraints from X-ray observations 
\citep{Ozel:2016oaf,Riley_2019,Miller_2019,Raaijmakers_2019} as well as 
gravitational wave data \citep{Abbott:2018exr}. They 
are also consistent with the most recent results inferred 
by the NICER experiment from observations of a rotation-powered pulsar 
\citep{Riley_2019,Miller_2019,Raaijmakers_2019}.

\section{Discussion and Conclusions}\label{Sec:conclusions}

Based on an analytic description of the spacetime around 
a NS in terms of three independent parameters (mass, 
spin, and quadrupole moment), we developed a novel method to 
calculate accurately the frequencies of geodesic motion in the
closest vicinity of the star, without resorting to a specific EoS, but 
instead adopting priors to our treatment that only allow specific 
parameter ranges for an unspecific EoS.
Once these frequencies are related to the
QPO frequencies observed in low mass X-ray binaries through a 
QPO model, they can be fit to the data in order to obtain estimates 
of the NS mass, spin and quadrupole moment, from which, in turn, 
an estimate of the NS radius and mass can be derived. This provides 
constraints to the EoS, and in principle can inform models of 
supranuclear density matter in a complementary way to other X-ray 
based techniques.

We presented a proof of principle of the method in application
to the observed pairs of kHz QPO frequencies (doublets) in three 
systems (\texttt{4U1608-52}, 
\texttt{4U0614+09} and \texttt{4U1728-34}) by applying the QPO
frequency identification proposed, for instance in the relativistic precession 
model \citep{Stella:1998mq}. 
Through a Bayesian analysis we obtained mass estimates around 
$\sim 2.05-2.16$~M$_{\odot}$, close to (but not exceeding) the maximum of 
the observed distribution of NS masses \citep{Lattimer:2019eez}.  

Our results appear to favour stiff EoSs, i.e. large NS masses and relatively large 
NS radii, as indicated by the marginalised distributions (see Fig. \ref{fig:massradius}). 
Our method yields relatively low values of the spin parameter $j$, all strictly smaller 
than 0.3. One may ask, how is it possible to produce any constrains on the parameter $\alpha$, 
since the geodesic frequencies used are such weak functions of that parameter. This is 
due to the fact that the spin $j$ and $\alpha$ also determine both the equatorial radius 
of the star and the ISCO. This means that the radius where the orbital motion occurs is 
limited by either the ISCO or the surface of the star, which limits the possible range of 
frequencies. These limitations in turn result in constraints on the parameters $j$ and $\alpha$. 

Once the values of $\alpha,j$ and $M$ are known, we can derive the probability 
distribution of the neutron star rotation rate $f$. To this aim we use the semi-analytic 
fit derived in \cite{Pappas:2015mba} (see eq.~B1 and the discussion in \citealt{Pappas:2015mba}). 
Due to the poor constraints on the quadrupole moment $\alpha$, the resulting 
bounds on $f$ are rather loose. We find that $f$ is constrained  at 90\% confidence 
interval within $\sim[0,582]$Hz ($f\sim[0,739]$Hz at 95\% confidence level) and $\sim[20,1000]$Hz 
($f\sim[18,1100]$Hz at 95\% confidence level) for \texttt{4U1608-52} 
and \texttt{4U0614+09}. $f$ is unconstrained for \texttt{4U1728-34}, for which we have a nearly flat distribution between 0 and 1000 Hz. These values are consistent (marginally, in the case of \texttt{4U1608-52}) with previous estimates of the NS spin frequencies for the above binary systems, i.e. $f\approx 620$Hz, $f\approx 415$Hz and $f\approx 363$Hz, respectively \citep{Watts2012}.
We stress that being this work a proof of concept, we decided to adopt priors with only weak limitations in order to assess the goodness of our approach. 
If additional information, chiefly the NS rotation period, were included as a prior in the analysis, significanlty tighter bounds would result. A precise determination of the NS mass, moment of inertia and quadrupole moment that can be obtained in this way would provide an unprecedented three-parameter constraint on the EoS, for which suitable 3D mappings of the EoS (as opposed to the familiar 2D mapping involving mass and radius, or mass and moment of inertia, \citealt{Lattimer:2006xb}) would be needed. Moreover, such measurements could be used together with current and future constraints form gravitational wave sources, to infer multi-messenger bounds on the stellar structure \citep{Fasano:2019zwm}.

\smallskip

The above mass and spin estimates are compatible with those from the 
early applications of the RPM, as the relevant geodesic frequencies 
depend only weakly on the spin and quadrupole moment (see, e.g., \citealt{Stella1998}, 
but also \citealt{duBuisson2019}). Instead, the EoS-independent bounds 
on the radius and mass that we derived represent original results of our 
new method. We note that the $\sim 2$~M$_{\odot}$ region in the NS 
mass-radius diagram is virtually unconstrained at present by observations 
in the radio and X-ray bands, while the limits on the tidal deformability from 
the GW170817 event, once translated into mass-radius bounds, are compatible 
with our results~\citep{Abbott:2018exr}. We remark that a precise measurement 
of the mass quadrupole moment can be used to directly constrain the EoS, 
without an explicit determination of the radius~\citep[see][]{Pappas:2012nt,Pappas:2013naa}. 
The limits on the quadrupole  that we obtained in this work are very loose due to the low number of QPOs considered (up to 8 for a given source)  and no stringent conclusion can be drawn. Instead, significantly tighter bounds are obtained by considering a larger number of doublets (Motta et al. in prep).

The method we presented is amenable to further, more extensive applications 
which exploit different QPO datasets and/or alternative QPO models. 
In a forthcoming study we will fit QPO {\it triplets}, each consisting of a low-frequency 
QPO simultaneous to the two kHz QPOs,  
where the former is associated with nodal precession frequency, in addition to the azimuthal and 
periastron precession frequencies that were used in the present study.
This is expected to yield higher precision estimates of 
the parameters governing the NS spacetime, since the nodal 
precession frequency depend quite strongly on 
NS spin and quadrupole moment. 
Our method is also applicable to 
other models relating QPOs to geodesic frequencies. This includes models 
in which QPO frequency identifications are different than those of the RPM, 
such as the epicyclic resonance models and global disk oscillation models in their various 
versions ({\it e.g.} \cite{Torok:2014eaa,2016A&A...586A.130S} and refs therein). In all cases a 
key requirement is that the set of independent QPO frequencies predicted by a model and fit to the datais sufficiently large that the three parameters describing NS spacetime can be derived.

The reliability of results from our new method, like that of inferences based on
QPOs in general, depend crucially on the correctness of the association of  
QPO signals to geodesic frequencies. Extensive, high sensitivity and large signal to noise 
QPO measurements to be obtained with next generation, large area X-ray missions, such 
as eXTP~\citep{DeRosa:2018aka}, Athena~\citep{Barcons:2015dua} and STROBE-X 
\citep{Ray:2019pxr} will allow the detection and precise characterisation of a significantly higher number of QPOs from many sources. In conjunction with detailed applications of competing models 
and advanced methods like our own, they may provide the key to resolving long-standing 
ambiguities in QPO interpretation and placing stringent constraints on NS structure
and EoS. 

\bigskip
\noindent{\em Acknowledgments.---}
We thank Tiziano Abdelsalhin for having carefully read 
this manuscript. 
AM, GP, and PP acknowledge financial support provided 
under the European Union's H2020 ERC, Starting Grant 
agreement no.~DarkGRA--757480.
AM acknowledges partial financial support provided under 
the European Union's H2020 ERC Consolidator Grant ``Matter 
and strong-field gravity: New frontiers in Einstein's 
theory'' grant agreement no. MaGRaTh--646597 during the 
development of this project. 
LS acknowledges financial contributions from ASI-INAF 
agreements 2017-14-H.O and  I/037/12/0 and 
from “iPeska” research grant (P.I. Andrea Possenti) funded 
under the INAF call PRIN-SKA/CTA (resolution 70/2016).
This project has received 
funding from the European Union's Horizon 2020 research 
and innovation programme under the Marie Sklodowska-Curie 
grant agreement No 690904.
SEM acknowledges the Violette and Samuel Glasstone Research Fellowship programme, and the Oxford Centre for Astrophysical Surveys, which is funded through generous support from the Hintze Family Charitable Foundation.
The authors would like to acknowledge networking support 
by the COST Action CA16104 and support from the Amaldi 
Research Center funded by the MIUR program ``Dipartimento 
di Eccellenza'' (CUP: B81I18001170001).

\appendix

\section{Metric components of rotating neutron stars}\label{app:metric}
In this appendix we present the explicit form of the 
metric functions introduced in Sec.~\ref{Sec:metric} for 
the line element given in Eq.~\eqref{Pap}, which describes 
the spacetime around a rotating NS. We refer the reader to 
\citet{Pappas:2016sye} for further details. The components 
of the metric are functions of the coordinates 
$(\rho,z)$, and of the multipole moments, and are given 
by the following expressions:

\beq   f(\rho,z) &=& 1-\frac{2 M}{\sqrt{\rho ^2+z^2}}+\frac{2 M^2}{\rho ^2+z^2}
                              +\frac{\left(M_2-M^3\right) \rho ^2-2 \left(M^3+M_2\right) z^2}{\left(\rho ^2+z^2\right)^{5/2}}  
                                    +\frac{2 z^2 \left(-J^2+M^4+2 M_2 M\right)-2 M M_2 \rho ^2}{\left(\rho ^2+z^2\right)^3}\nn\\
                                    && +\frac{A(\rho,z)}{28 \left(\rho ^2+z^2\right)^{9/2}}+\frac{B(\rho,z)}{14 \left(\rho ^2+z^2\right)^5}\ ,\label{ref:f}\\
 \omega(\rho,z)  &=& -\frac{2 J \rho ^2}{\left(\rho ^2+z^2\right)^{3/2}}   -\frac{2 J M \rho ^2}{\left(\rho ^2+z^2\right)^2}  +\frac{F(\rho,z)}{\left(\rho ^2+z^2\right)^{7/2}}+\frac{H(\rho,z)}{2 \left(\rho ^2+z^2\right)^4}+ \frac{G(\rho,z)}{4 \left(\rho ^2+z^2\right)^{11/2}}\ ,\label{ref:omega}\\                             
               \zeta(\rho,z) &=& \frac{\rho ^2 \left[J^2 \left(\rho ^2-8 z^2\right)+M \left(M^3+3 M_2\right) \left(\rho ^2-4 z^2\right)\right]}{4 \left(\rho
   ^2+z^2\right)^4} -\frac{M^2 \rho ^2}{2 \left(\rho ^2+z^2\right)^2}\ ,   \label{ref:gamma}                  
                                    \eeq
where
\beq   A(\rho,z)  &=&   8 \rho ^2 z^2 \left(24 J^2 M+17 M^2 M_2+21 M_4\right)
      +\rho ^4 \left(-10 J^2 M+7 M^5+32 M_2 M^2-21 M_4\right)\nn\\
      && +8 z^4 \left(20 J^2 M-7 M^5-22 M_2 M^2-7 M_4\right)\ , \\
   B(\rho,z)   &=&   \rho ^4 \left(10 J^2 M^2+10 M_2 M^3+21 M_4M+7 M_2^2\right)
         +4 z^4 \left(-40 J^2 M^2-14 J S_3+7 M^6+30 M_2 M^3 \right.\nn\\
          && \left.+14 M_4 M+7 M_2^2\right)-4 \rho ^2 z^2 \left(27 J^2 M^2-21 J S_3+7 M^6+48 M_2 M^3+42 M_4 M+7 M_2^2\right)\ ,\\
 H(\rho,z)  &=&  4 \rho ^2 z^2 \left[J \left(M_2-2 M^3\right)-3 M S_3\right]+\rho ^4 \left(J M_2+3 M S_3\right)\ , \\
   G(\rho,z)  &=& \rho ^2 \left\{-J^3\left(\rho ^4+8 z^4-12 \rho ^2 z^2\right)+J M \left[\left(M^3+2 M_2\right) \rho ^4-8 \left(3 M^3+2 M_2\right) z^4\right.\right.\nn\\
                               && \left.\left.+4 \left(M^3+10 M_2\right) \rho ^2 z^2\right]+M^2 S_3 \left(3 \rho ^4-40 z^4+12 \rho ^2 z^2\right)\right\}\ , \\
   F(\rho,z)  &=&  \rho ^4 \left(S_3-J M^2\right)-4 \rho ^2 z^2 \left(J  M^2+S_3\right)\ .   \eeq

   The spacetime defined above can be given in an even 
   more convenient form so as to have the right Schwarzschild 
   limit when the rotation goes to zero, i.e., $j\rightarrow0$.
   To this aim, we resum the expansions of $f(\rho,z)$ and
   $\zeta(\rho,z)$, using the variable $r=\sqrt{\rho^2+z^2}$, 
   such that when the rotation vanishes the metric coincides
   with the exact Schwarzschild solution in its standard form.  With this procedure,
   we obtain:
 \beq f(\rho,z) &=& 1-\frac{4 M}{r_- + r_+ +2 M} + \frac{\alpha ^2 j^4 M^6 \left(\rho ^2-2 z^2\right)^2}{2 r^{10}} 
+\frac{2 \beta  j^4 M^6 z^2  \left(2 z^2-3 \rho ^2\right)}{r^{10}}\nn\\
&-&\frac{\gamma  j^4 M^5 \left(3 \rho ^4+8 z^4-24 \rho ^2 z^2\right) \left(r-2 M \right)}{4 r^{10}}
-\frac{j^2 M^4 }{14 r^{10}} \left[2 M^2 \left(-5 \rho ^4+80 z^4+54 \rho ^2 z^2\right) \right.\nn\\
&-&\left.M \left(20 z^2-\rho ^2\right) \left(5 \rho ^2+4 z^2\right) r+28 z^2  r^4\right] +\frac{\alpha  j^2 M^3 }{7 r^{10}}\left[M^3  \left(-5 \rho ^4-60 z^4+96 \rho ^2 z^2\right)\right.\nn\\
&+&2 M^2 r \left(-4 \rho ^4+22 z^4-17 \rho ^2 z^2\right)+\left.\!14 M \left(\rho ^6-2 z^6-3 \rho ^2 z^4\right)+7 \left(2 z^2-\rho ^2\right)  r^5 \right]
    \eeq
and
\begin{equation}
 \zeta(\rho,z)= \frac{1}{2} \log \left(\frac{r^2-M^2+r_- r_+ }{2 r_- r_+}\right) + \frac{j^2 M^4 \rho ^2 \left[(1-3 \alpha ) \rho ^2+4 (3 \alpha -2) z^2\right]}{4 r^8}\ , 
\end{equation}
where $r_{\pm}=\sqrt{(M\pm z)^2+\rho ^2}$ and we have used 
the definitions \eqref{multipole} for the moments. When $j 
\rightarrow 0$, from Eq.~\eqref{ref:omega} it follows that 
$\omega \rightarrow 0$ and the functions $f$ and $\zeta$ 
take their Schwarzschild form. This metric, as the previous 
one, is accurate up to $M_4$ in the moments and up to order 
$\mathcal{O}(M^6/r^6)$ with respect to the vacuum field 
equations. 
    
It is worth remarking that while the spacetime is given in 
Weyl-Papapetrou coordinates, which differ from the usual 
Schwarzschild-like or quasi-isotropic ones, the various 
geodesic frequencies are coordinate-independent quantities, 
while the relevant radii on the equatorial plane can be 
expressed in terms of the circumferential radius which is 
also a geometric and coordinate-independent quantity.   

\section{QPO frequencies}\label{app:freqs}

In this appendix we give the QPO frequencies and
corresponding uncertainties for the three LMXB 
systems considered in this paper. Frequencies are all from 
RXTE/PCA observations and were taken from \cite{vanDoesburgh2017}.


\begin{table}[ht]
  \centering
  \vspace{0.2cm}
\texttt{4U1608-52}
  \vspace{0.2cm}\\
    \begin{tabular}{ccccccc}
   \hline
   \hline
 Doublet \# & $\nu _{\phi }$ & $\sigma ^{(+)}_{\phi }$ & $\sigma ^{(-)}_{\phi }$&  $\nu _{\text{per}}$ & $\sigma ^{(+)}_{\text{per}}$ &
 $\sigma^{(-)}_{\text{per}}$ \\
 1 & 849.92 & 6.94 & 6.53 & 535.32 & 15.4 & 23.1 \\
 2 & 940.93 & 12.1 & 12.5 & 655.78 & 2.15 & 2.07  \\
 3 & 958.61 & 8.19 & 8.36 & 654.7 & 0.23 & 0.23  \\
 4 & 976.6 & 6.89 & 7.00. & 674.76 & 1.26 & 1.24 \\
 5 & 1034.6 & 10.6 & 10.3 & 769.32 & 0.83 & 0.79  \\
 6 & 1041.1 & 7.04 & 7.32 & 774.82 & 0.83 & 0.81 \\
 7 & 1053.1 & 11.2 & 13.6 & 740.61 & 0.59 & 0.54 \\
   \hline
  \end{tabular}  \\
  \vspace{0.2cm}
\texttt{4U0614+09}
  \vspace{0.2cm}\\
  \begin{tabular}{ccccccc}
   \hline
   \hline
 \# & $\nu _{\phi }$ & $\sigma ^{(+)}_{\phi }$ & $\sigma ^{(-)}_{\phi }$&  $\nu _{\text{per}}$ & $\sigma ^{(+)}_{\text{per}}$ &
 $\sigma^{(-)}_{\text{per}}$ \\
 1 &  957.11 & 8.97 & 9.24 & 636.61 & 1.98 & 2.1 \\
 2 &  959.41 & 7.06 & 7.73 & 649.9 & 1.61 & 1.8  \\
 3 & 1076.4 & 11.2 & 14.4 & 749.84 & 1.77 & 1.68  \\
 4 & 1103.8 & 10.7 & 11.1 & 761.02 & 1.21 & 1.29 \\
 5 &  1166.7 & 16.9 & 21.7 & 753.15 & 5.67 & 5.23 \\   \hline
   \hline
  \end{tabular}  
\\
  \vspace{0.2cm}
\texttt{4U1728-34}
  \vspace{0.2cm}\\
  \begin{tabular}{ccccccc}
   \hline
   \hline
 \# & $\nu _{\phi }$ & $\sigma ^{(+)}_{\phi }$ & $\sigma ^{(-)}_{\phi }$&  $\nu _{\text{per}}$ & $\sigma ^{(+)}_{\text{per}}$ &
 $\sigma^{(-)}_{\text{per}}$\\
 1 & 717.9 & 5.09 & 5.04 & 377. & 18.6 & 15. \\
 2 & 873.25 & 3.36 & 3.3 & 538.38 & 37.4 & 37.1 \\
 3 & 972.49 & 5.68 & 5.51 & 614.15 & 3.66 & 4.2 \\
 4 & 1089.2 & 3.85 & 3.97 & 752.42 & 0.67 & 0.66 \\
 5 & 1091.4 & 10.6 & 10.8 & 740.48 & 0.84 & 0.87  \\
 6 & 1107.3 & 9.99 & 9.72 & 778.22 & 2.85 & 2.64 \\
 7 & 1118.8 & 7.29 & 7.53 & 801.78 & 10.8 & 11.  \\
 8 & 1149.9 & 1.58 & 1.16 & 816.36 & 1.08 & 1.21  \\   
 \hline
   \hline
  \end{tabular}  
  \caption{QPO frequencies (with experimental errors $\sigma^{(\pm)}$) observed for the three sources analysed in this 
  paper, \texttt{4U1608-52}, \texttt{4U0614+09}, and \texttt{4U1728-34}. According the RPM, $(\nu_\phi,\nu_\tn{per})$ 
  correspond to the kHz QPO doublets.}\label{table_freqs}
\end{table}

\bibliography{bibnote}

\begin{thebibliography}{}
\expandafter\ifx\csname natexlab\endcsname\relax\def\natexlab#1{#1}\fi

\bibitem[{Abbott {et~al.}(2018)}]{Abbott:2018exr}
Abbott, B.~P., {et~al.} 2018, Phys. Rev. Lett., 121, 161101

\bibitem[{Abbott {et~al.}(2019)}]{Abbott:2018wiz}
---. 2019, Phys. Rev., X9, 011001

\bibitem[{Abramowicz \& Kluzniak(2001)}]{Abramowicz:2001bi}
Abramowicz, M.~A., \& Kluzniak, W. 2001, Astron. Astrophys., 374, L19

\bibitem[{Akmal {et~al.}(1998)Akmal, Pandharipande, \&
  Ravenhall}]{Akmal:1998cf}
Akmal, A., Pandharipande, V.~R., \& Ravenhall, D.~G. 1998, Phys. Rev., C58,
  1804

\bibitem[{Armitage \& Natarajan(1999)}]{Armitage:1999aj}
Armitage, P.~J., \& Natarajan, P. 1999, Astrophys. J., 525, 909

\bibitem[{Barcons {et~al.}(2015)Barcons, Nandra, Barret, den Herder, Fabian,
  Piro, \& Watson}]{Barcons:2015dua}
Barcons, X., Nandra, K., Barret, D., {et~al.} 2015, J. Phys. Conf. Ser., 610,
  012008

\bibitem[{Belloni \& Stella(2014)}]{Belloni:2014iqa}
Belloni, T.~M., \& Stella, L. 2014, Space Sci. Rev., 183, 43

\bibitem[{De~Rosa {et~al.}(2019)}]{DeRosa:2018aka}
De~Rosa, A., {et~al.} 2019, Sci. China Phys. Mech. Astron., 62, 29504

\bibitem[{Doneva \& Pappas(2018)}]{DDGP2017arXiv}
Doneva, D.~D., \& Pappas, G. 2018, Universal Relations and Alternative Gravity
  Theories, ed. L.~Rezzolla, P.~Pizzochero, D.~I. Jones, N.~Rea, \&
  I.~Vida{\~{n}}a (Cham: Springer International Publishing), 737--806

\bibitem[{Douchin \& Haensel(2001)}]{Douchin:2001sv}
Douchin, F., \& Haensel, P. 2001, Astron. Astrophys., 380, 151

\bibitem[{{du Buisson} {et~al.}(2019){du Buisson}, {Motta}, \&
  {Fender}}]{duBuisson2019}
{du Buisson}, L., {Motta}, S., \& {Fender}, R. 2019, \mnras, 486, 4485

\bibitem[{Fasano {et~al.}(2019)Fasano, Abdelsalhin, Maselli, \&
  Ferrari}]{Fasano:2019zwm}
Fasano, M., Abdelsalhin, T., Maselli, A., \& Ferrari, V. 2019, Phys. Rev.
  Lett., 123, 141101

\bibitem[{Fodor {et~al.}(1989)Fodor, Hoenselaers, \& Perj\'{e}s}]{fodor:2252}
Fodor, G., Hoenselaers, C., \& Perj\'{e}s, Z. 1989, J.Math.Phys., 30, 2252

\bibitem[{{Fragile} {et~al.}(2016){Fragile}, {Straub}, \&
  {Blaes}}]{Fragile2016}
{Fragile}, P.~C., {Straub}, O., \& {Blaes}, O. 2016, \mnras, 461, 1356

\bibitem[{Geroch(1970{\natexlab{a}})}]{Geroch70I}
Geroch, R.~P. 1970{\natexlab{a}}, J.Math.Phys., 11, 1955

\bibitem[{Geroch(1970{\natexlab{b}})}]{Geroch70II}
---. 1970{\natexlab{b}}, J.Math.Phys., 11, 2580

\bibitem[{Gilks {et~al.}(1996)Gilks, Richardson, \& Spiegelhalter}]{Gilks:1996}
Gilks, W.~R., Richardson, S., \& Spiegelhalter, D.~J. 1996, {Markov Chain Monte
  Carlo in Practice} (London, UK: Chapman \& Hall)

\bibitem[{Hansen(1974)}]{hansen:46}
Hansen, R.~O. 1974, J. Math. Phys., 15, 46

\bibitem[{Hinderer {et~al.}(2018)Hinderer, Rezzolla, \&
  Baiotti}]{Hinderer:2018mrj}
Hinderer, T., Rezzolla, L., \& Baiotti, L. 2018, Astrophys. Space Sci. Libr.,
  457, 575

\bibitem[{{Kato}(2001)}]{2001PASJ...53L..37K}
{Kato}, S. 2001, Publications of the Astronomical Society of Japan, 53, L37

\bibitem[{Kato(2004)}]{Kato:2004vs}
Kato, S. 2004, Publ. Astron. Soc. Jap., 56, 905

\bibitem[{Kato(2012)}]{Kato:2012mb}
---. 2012, Publ. Astron. Soc. Jap., 64, 139

\bibitem[{kato(2012)}]{kato:2012hq}
kato, S. 2012, Publ. Astron. Soc. Jap., 64, 129

\bibitem[{Kluzniak \& Abramowicz(2001)}]{Kluzniak:2001ar}
Kluzniak, W., \& Abramowicz, M.~A. 2001, arXiv:astro-ph/0105057

\bibitem[{Kluzniak \& Abramowicz(2002)}]{Kluzniak:2002bb}
---. 2002, arXiv:astro-ph/0203314

\bibitem[{Lattimer(2019)}]{Lattimer:2019eez}
Lattimer, J.~M. 2019, Universe, 5, 159

\bibitem[{Lattimer \& Prakash(2007)}]{Lattimer:2006xb}
Lattimer, J.~M., \& Prakash, M. 2007, Phys. Rept., 442, 109

\bibitem[{Markovic \& Lamb(1998)}]{Markovic:1998qy}
Markovic, D., \& Lamb, F.~K. 1998, arXiv:astro-ph/9801075

\bibitem[{Maselli {et~al.}(2015{\natexlab{a}})Maselli, Gualtieri, Pani, Stella,
  \& Ferrari}]{Maselli:2014fca}
Maselli, A., Gualtieri, L., Pani, P., Stella, L., \& Ferrari, V.
  2015{\natexlab{a}}, Astrophys.J., 801, 115

\bibitem[{Maselli {et~al.}(2015{\natexlab{b}})Maselli, Pani, Gualtieri, \&
  Ferrari}]{Maselli:2015tta}
Maselli, A., Pani, P., Gualtieri, L., \& Ferrari, V. 2015{\natexlab{b}}, Phys.
  Rev., D92, 083014

\bibitem[{Miller {et~al.}(2019{\natexlab{a}})}]{Miller:2019cac}
Miller, M.~C., {et~al.} 2019{\natexlab{a}}, Astrophys. J. Lett., 887, L24

\bibitem[{Miller {et~al.}(2019{\natexlab{b}})Miller, Lamb, Dittmann, Bogdanov,
  Arzoumanian, Gendreau, Guillot, Harding, Ho, Lattimer, Ludlam, Mahmoodifar,
  Morsink, Ray, Strohmayer, Wood, Enoto, Foster, Okajima, Prigozhin, \&
  Soong}]{Miller_2019}
Miller, M.~C., Lamb, F.~K., Dittmann, A.~J., {et~al.} 2019{\natexlab{b}}, The
  Astrophysical Journal, 887, L24

\bibitem[{{Morsink} \& {Stella}(1999)}]{1999ApJ...513..827M}
{Morsink}, S.~M., \& {Stella}, L. 1999, \apj, 513, 827

\bibitem[{{Motta} {et~al.}(2014{\natexlab{a}}){Motta}, {Belloni}, {Stella},
  {Mu{\~n}oz-Darias}, \& {Fender}}]{Motta2014}
{Motta}, S.~E., {Belloni}, T.~M., {Stella}, L., {Mu{\~n}oz-Darias}, T., \&
  {Fender}, R. 2014{\natexlab{a}}, \mnras, 437, 2554

\bibitem[{{Motta} {et~al.}(2014{\natexlab{b}}){Motta}, {Munoz-Darias}, {Sanna},
  {Fender}, {Belloni}, \& {Stella}}]{2014MNRAS.439L..65M}
{Motta}, S.~E., {Munoz-Darias}, T., {Sanna}, A., {et~al.} 2014{\natexlab{b}},
  \mnras, 439, L65

\bibitem[{M\"uller \& Sbalzarini(2010)}]{5586491}
M\"uller, C.~L., \& Sbalzarini, I.~F. 2010, IEEE Congress on Evolutionary
  Computation, 1

\bibitem[{{Nowak} \& {Wagoner}(1991)}]{1991ApJ...378..656N}
{Nowak}, M.~A., \& {Wagoner}, R.~V. 1991, Astrophys. J., 378, 656

\bibitem[{{Nowak} \& {Wagoner}(1992)}]{1992ApJ...393..697N}
---. 1992, Astrophys. J., 393, 697

\bibitem[{Nowak {et~al.}(1997)Nowak, Wagoner, Begelman, \& Lehr}]{Nowak:1996hg}
Nowak, M.~A., Wagoner, R.~V., Begelman, M.~C., \& Lehr, D.~E. 1997, Astrophys.
  J., 477, L91

\bibitem[{Ozel \& Freire(2016)}]{Ozel:2016oaf}
Ozel, F., \& Freire, P. 2016, Ann. Rev. Astron. Astrophys., 54, 401

\bibitem[{{{\"O}zel} {et~al.}(2012){{\"O}zel}, {Psaltis}, {Narayan}, \& {Santos
  Villarreal}}]{Ozel2012}
{{\"O}zel}, F., {Psaltis}, D., {Narayan}, R., \& {Santos Villarreal}, A. 2012,
  \apj, 757, 55

\bibitem[{{Papapetrou}(1953)}]{Papapetrou}
{Papapetrou}, A. 1953, Annalen der Physik, 447, 309

\bibitem[{Pappas(2012)}]{Pappas:2012nt}
Pappas, G. 2012, Mon. Not. Roy. Astron. Soc., 422, 2581

\bibitem[{Pappas(2015)}]{Pappas:2015mba}
---. 2015, Mon. Not. Roy. Astron. Soc., 454, 4066

\bibitem[{Pappas(2017)}]{Pappas:2016sye}
---. 2017, Mon. Not. Roy. Astron. Soc., 466, 4381

\bibitem[{Pappas \& Apostolatos(2013)}]{Pappas:2012nv}
Pappas, G., \& Apostolatos, T.~A. 2013, Mon. Not. Roy. Astron. Soc., 429, 3007

\bibitem[{Pappas \& Apostolatos(2014)}]{Pappas:2013naa}
---. 2014, Phys. Rev. Lett., 112, 121101

\bibitem[{Raaijmakers {et~al.}(2019)Raaijmakers, Riley, Watts, Greif, Morsink,
  Hebeler, Schwenk, Hinderer, Nissanke, Guillot, Arzoumanian, Bogdanov,
  Chakrabarty, Gendreau, Ho, Lattimer, Ludlam, \& Wolff}]{Raaijmakers_2019}
Raaijmakers, G., Riley, T.~E., Watts, A.~L., {et~al.} 2019, The Astrophysical
  Journal, 887, L22

\bibitem[{Ray {et~al.}(2019)}]{Ray:2019pxr}
Ray, P.~S., {et~al.} 2019, arXiv:1903.03035

\bibitem[{Riley {et~al.}(2019{\natexlab{a}})}]{Riley:2019yda}
Riley, T.~E., {et~al.} 2019{\natexlab{a}}, Astrophys. J. Lett., 887, L21

\bibitem[{Riley {et~al.}(2019{\natexlab{b}})Riley, Watts, Bogdanov, Ray,
  Ludlam, Guillot, Arzoumanian, Baker, Bilous, Chakrabarty, Gendreau, Harding,
  Ho, Lattimer, Morsink, \& Strohmayer}]{Riley_2019}
Riley, T.~E., Watts, A.~L., Bogdanov, S., {et~al.} 2019{\natexlab{b}}, The
  Astrophysical Journal, 887, L21

\bibitem[{Stein {et~al.}(2014)Stein, Yagi, \& Yunes}]{Stein:2013ofa}
Stein, L.~C., Yagi, K., \& Yunes, N. 2014, Astrophys. J., 788, 15

\bibitem[{Stella \& Vietri(1998)}]{Stella:1997tc}
Stella, L., \& Vietri, M. 1998, Astrophys. J., 492, L59

\bibitem[{{Stella} \& {Vietri}(1998)}]{Stella1998}
{Stella}, L., \& {Vietri}, M. 1998, \apjl, 492, L59+

\bibitem[{Stella \& Vietri(1999)}]{Stella:1998mq}
Stella, L., \& Vietri, M. 1999, Phys. Rev. Lett., 82, 17

\bibitem[{Stella {et~al.}(1999)Stella, Vietri, \& Morsink}]{Stella:1999sj}
Stella, L., Vietri, M., \& Morsink, S. 1999, Astrophys. J., 524, L63

\bibitem[{Strohmayer {et~al.}(1996)Strohmayer, Zhang, Swank, Smale, Titarchuk,
  Day, \& Lee}]{Strohmayer:1996zz}
Strohmayer, T.~E., Zhang, W., Swank, J.~H., {et~al.} 1996, Astrophys. J., 469,
  L9

\bibitem[{{Stuchl{\'\i}k} \& {Kolo{\v{s}}}(2016)}]{2016A&A...586A.130S}
{Stuchl{\'\i}k}, Z., \& {Kolo{\v{s}}}, M. 2016, \aap, 586, A130

\bibitem[{Swank {et~al.}(1994)}]{Swank:1994zg}
Swank, J.~H., {et~al.} 1994

\bibitem[{{Syunyaev}(1973)}]{1973SvA....16..941S}
{Syunyaev}, R.~A. 1973, \sovast, 16, 941

\bibitem[{{T{\"o}r{\"o}k} {et~al.}(2005){T{\"o}r{\"o}k}, {Abramowicz},
  {Klu{\'z}niak}, \& {Stuchl{\'\i}k}}]{Torok2005}
{T{\"o}r{\"o}k}, G., {Abramowicz}, M.~A., {Klu{\'z}niak}, W., \&
  {Stuchl{\'\i}k}, Z. 2005, \aap, 436, 1

\bibitem[{Torok {et~al.}(2014)Torok, Bakala, Sramkova, Stuchlik, Urbanec, \&
  Goluchova}]{Torok:2014eaa}
Torok, G., Bakala, P., Sramkova, E., {et~al.} 2014, arXiv:1408.4220,
  [Astrophys. J.760,138(2012)]

\bibitem[{Tsang \& Pappas(2016)}]{Tsang:2015bty}
Tsang, D., \& Pappas, G. 2016, Astrophys. J., 818, L11

\bibitem[{Urbancova {et~al.}(2019)Urbancova, Urbanec, Torok, Stuchlik,
  Blaschke, \& Miller}]{Urbancova:2019btk}
Urbancova, G., Urbanec, M., Torok, G., {et~al.} 2019, arXiv:1905.00730

\bibitem[{{van der Klis}(1995)}]{vanderKlis1995a}
{van der Klis}, M. 1995, IAU Colloq. 151: Flares and Flashes, 454, 321

\bibitem[{{van der Klis}(2006)}]{2006csxs.book...39V}
---. 2006, {Rapid X-ray Variability}, Vol.~39, 39--112

\bibitem[{{van der Klis} {et~al.}(1996){van der Klis}, {Swank}, {Zhang},
  {Jahoda}, {Morgan}, {Lewin}, {Vaughan}, \& {van Paradijs}}]{vanderKlis1996}
{van der Klis}, M., {Swank}, J.~H., {Zhang}, W., {et~al.} 1996, \apjl, 469, L1

\bibitem[{{van Doesburgh} \& {van der Klis}(2017)}]{vanDoesburgh2017}
{van Doesburgh}, M., \& {van der Klis}, M. 2017, \mnras, 465, 3581

\bibitem[{Wagoner {et~al.}(2001)Wagoner, Silbergleit, \&
  Ortega-Rodriguez}]{Wagoner:2001uj}
Wagoner, R.~V., Silbergleit, A.~S., \& Ortega-Rodriguez, M. 2001, Astrophys.
  J., 559, L25

\bibitem[{{Watts}(2012)}]{Watts2012}
{Watts}, A.~L. 2012, \araa, 50, 609

\bibitem[{Watts {et~al.}(2016)}]{Watts:2016uzu}
Watts, A.~L., {et~al.} 2016, Rev. Mod. Phys., 88, 021001

\bibitem[{Wiringa {et~al.}(1988)Wiringa, Fiks, \& Fabrocini}]{PhysRevC.38.1010}
Wiringa, R.~B., Fiks, V., \& Fabrocini, A. 1988, Phys. Rev. C, 38, 1010

\bibitem[{Yagi {et~al.}(2014)Yagi, Kyutoku, Pappas, Yunes, \&
  Apostolatos}]{Yagi:2014bxa}
Yagi, K., Kyutoku, K., Pappas, G., Yunes, N., \& Apostolatos, T.~A. 2014, Phys.
  Rev., D89, 124013

\bibitem[{Yagi \& Yunes(2017)}]{Yagi:2016bkt}
Yagi, K., \& Yunes, N. 2017, Phys. Rept., 681, 1

\end{thebibliography}
\bibliographystyle{apj}

\end{document}